\global\def\draftcontrol{0}

   \def\versionno{cfttransport}

\catcode`\@=11

\expandafter\ifx\csname draftcontrol\endcsname\relax\global\def\draftcontrol{0}
\fi

{\count255=\time\divide\count255 by 60
\xdef\hourmin{\number\count255}
\multiply\count255 by-60\advance\count255 by\time
\xdef\hourmin{\hourmin:\ifnum\count255<10 0\fi\the\count255}}
\def\draftdate{\number\month/\number\day/\number\year\ \ \ \hourmin }

\newcommand\makepapertitle{\par
  \begingroup
    \renewcommand\thefootnote{\@fnsymbol\c@footnote}%
    \def\@makefnmark{\rlap{\@textsuperscript{\normalfont\@thefnmark}}}%
    \long\def\@makefntext##1{\parindent 1em\noindent
            \hb@xt@1.8em{%
                \hss\@textsuperscript{\normalfont\@thefnmark}}##1}%
     \newpage
     \global\@topnum\z@   
     \@makepapertitle
     \thispagestyle{empty}\@thanks
  \endgroup
  \setcounter{footnote}{0}%
  \global\let\thanks\relax
  \global\let\makepapertitle\relax
  \global\let\@makepapertitle\relax
  \global\let\@thanks\@empty
  \global\let\@author\@empty
  \global\let\@date\@empty
  \global\let\@title\@empty
  \global\let\title\relax
  \global\let\author\relax
  \global\let\date\relax
  \global\let\and\relax
  \def\version{\let\version\@version\@gobble}
}
\def\@makepapertitle{%
  \newpage
   \ifnum\draftcontrol=1 {}
   \version\versionno
   \vskip 3em%
   \else
   \hfill\hbox to 3cm {\parbox{4cm}{\@pubnum}\hss}%
   \vskip 3em%
   \fi
   \begin{center}%
   \let \footnote \thanks
     {\LARGE {\@title}}%
     \vskip 1.5em%
     {\normalsize
       \lineskip .5em%
       \begin{tabular}[t]{c}%
         \@author
       \end{tabular}\par}%
     \vskip 1.5em%
     {\@bstract}%
     \end{center}%
     \vskip 1.5em
     \@date%
   \par
}

\gdef\@pubnum{}
\def\pubnum#1{%
  \gdef\@pubnum{#1}}

\gdef\@bstract{}
\def\Abstract#1{%
  \gdef\@bstract{%
   \parbox{\textwidth-0pc}{%
   \centerline{\bf Abstract}\penalty1000%
\kern.2cm%
\noindent
\renewcommand\baselinestretch{1.0}%
{#1}}}
}

\def\ps@paper{\let\@mkboth\@gobbletwo%
     \ifnum\draftcontrol=1
    \def\@oddfoot{\hbox to \textwidth{\tiny \versionno \hfil\tiny\draftdate}%
    \hskip -\textwidth \hbox to \textwidth{\hfil\rm\thepage\hfil}}%
     \else\def\@oddfoot{\hbox to \textwidth{\hfil\rm\thepage\hfil}}
     \fi
     \let\@evenfoot\@oddfoot
}

\def\body{\clearpage
          \pagestyle{paper}
    }

\def\@version#1{\ifnum\draftcontrol=1
\typeout{}\typeout{#1}\typeout{}
\vskip3mm\centerline{\hbox{\fbox{\normalsize{\tt DRAFT -- #1 -- }
                   {\draftdate}}}}\vskip3mm
\fi}
\let\version\@version
\long\def\eqlabel#1{\ifnum\draftcontrol=1
                    \tag@false  
                    \tag*{(\theequation) \hbox to -0.2cm{\hspace{0cm}\small{#1}\hss}}
                    \refstepcounter{equation}
                    \edef\@currentlabel{\theequation}
                    \ltx@label{#1}          
                    \else
                    \label{#1}
                    \fi
                    }
\let\st@bibitem\@bibitem
\let\st@lbibitem\@lbibitem
\ifnum\draftcontrol=1
  \def\@bibitem#1{%
    \st@bibitem{#1}\a@@label{#1}\ignorespaces}
  \def\@lbibitem[#1]#2{%
    \st@lbibitem[#1]{#2}\a@@label{#2}\ignorespaces}
  \def\a@@label#1{%
    \gdef\a@lab{\smash{\normalfont\small#1}}
    \ifvmode
      \if@inlabel
        \global\setbox\@labels\hbox{%
          \llap{\a@lab\let\a@lab\relax
                \kern\@totalleftmargin\kern\marginparsep}%
          \box\@labels}%
      \fi
    \fi}
\fi

\documentclass[12pt,letterpaper]{article}

\usepackage{amsmath,amssymb,array,calc,epsfig,rotating,bm}
\usepackage[sort]{cite}
\usepackage{graphicx}
\usepackage{psfrag,verbatim}
\usepackage{xcolor}

\ifnum\draftcontrol=1
\fi

\renewcommand\baselinestretch{1.25}
\renewcommand\section{\@startsection {section}{1}{\z@}%
                                   {-3.5ex \@plus -1ex \@minus -.2ex}%
                                   {2.3ex \@plus.2ex}%
                                   {\normalfont\large\bfseries}}
\renewcommand\subsection{\@startsection{subsection}{2}{\z@}%
                                   {-3.25ex\@plus -1ex \@minus -.2ex}%
                                   {1.5ex \@plus .2ex}%
                                   {\normalfont\normalsize\bfseries}}
\renewcommand\subsubsection{\@startsection{subsubsection}{3}{\z@}%
                                   {-3.25ex\@plus -1ex \@minus -.2ex}%
                                   {1.5ex \@plus .2ex}%
                                   {\normalfont\normalsize\it}}
\renewcommand\paragraph{\@startsection{paragraph}{4}{\z@}%
                                   {-3.25ex\@plus -1ex \@minus -.2ex}%
                                   {1.5ex \@plus .2ex}%
                                   {\normalfont\normalsize\bf}}
      \newcommand{\hred}[1]{%
  \colorbox{red!50}{$\displaystyle#1$}}

\def\revise#1       {\raisebox{-0em}{\rule{3pt}{1em}}%
                     \marginpar{\raisebox{.5em}{\vrule width3pt\
                     \vrule width0pt height 0pt depth0.5em
                     \hbox to 0cm{\hspace{0cm}{%
                     \parbox[t]{4em}{\raggedright\footnotesize{#1}}}\hss}}}}

\newcommand\nxt[1]  {\\\fnxt#1}
\newcommand{\ie}{{\it i.e.,}\ }
\newcommand{\eg}{{\it e.g.,}\ }

\def\cala         {{\cal A}}

\def\calb         {{\cal B}}
\def\calc         {{\cal C}}
\def\cald         {{\cal D}}
\def\cale         {{\cal E}}
\def\calf         {{\cal F}}

\def\calo         {{\cal O}}

\def\zet          {{\mathbb Z}}

\def\del          {\partial}

\def\Re           {{\rm Re\hskip0.1em}}
\def\Im           {{\rm Im\hskip0.1em}}

\def\sqr#1#2{{\vcenter{\vbox{\hrule height.#2pt
 \hbox{\vrule width.#2pt height#1pt \kern#1pt
 \vrule width.#2pt}\hrule height.#2pt}}}}

\newcommand{\ft}[2]{{\textstyle{\frac{#1}{#2}}}}

\newcommand{\kk}{\mathfrak{q}}
\newcommand{\ww}{\mathfrak{w}}

\def\tV {{\tilde{V}}}
\def\tdV {{\del \tV}}
\def\tk {{(\alpha\kk)}}
\def\tw {{(\alpha\ww)}}

\def\hx{\hat{x}}

\def\aa1{\phi}
\def\cc1{\psi}

\catcode`\@=12

\begin{document}


\title{\bf The fate of the conformal order}

\date{November 23, 2020}

\author{
Alex Buchel \\
\it Department of Applied Mathematics\\
\it Department of Physics and Astronomy\\ 
\it University of Western Ontario\\
\it London, Ontario N6A 5B7, Canada\\
\it Perimeter Institute for Theoretical Physics\\
\it Waterloo, Ontario N2J 2W9, Canada\\[0.4cm]
}

\Abstract{We use holographic correspondence to study transport
of the conformal plasma in ${\mathbb R}^{2,1}$ in a phase with
a spontaneously broken global $\zet_2$ symmetry. The dual black branes in a
Poincare patch of asymptotically $AdS_4$ have ``hair'' ---
a condensate of the order parameter for the broken symmetry.
This hair affects both the hydrodynamic and the nonhydrodynamic quasinormal
modes of the black branes. Nonetheless, the shear viscosity of the conformal order
is universal, the bulk viscosity vanishes and the speed of the sound waves is $c_s^2=\ft 12$.
We compute the low-lying spectrum of the non-hydrodynamic modes. We identify a quasinormal mode
associated with the fluctuations of the $\zet_2$ order parameter with the positive imaginary part.
The presence of this mode in the spectrum renders the holographic conformal order
perturbatively unstable. Correspondingly, the dual black branes violate the correlated stability
conjecture.
}

\makepapertitle

\body

\version\versionno
\tableofcontents

\section{Introduction and summary}\label{intro}

Following the general suggestion of \cite{Chai:2020zgq}, we proposed in \cite{Buchel:2020thm}\footnote{See also
\cite{Buchel:2020xdk}.}
a holographic model for a conformal order\footnote{See \cite{Buchel:2009ge} for related nonconformal models.}:
a thermal phase of a conformal gauge theory in ${\mathbb R}^{2,1}$ with a nonzero  expectation value of an
irrelevant, dimension $\Delta=4$   operator $\calo$, spontaneously breaking the global $\zet_2$
symmetry. Specifically, for a  model $S_{{CFT_3^\psi}}$ in \cite{Buchel:2020thm}
two distinct thermal phases were identified:
\begin{equation}
\frac{\hat{\calf}}{(\pi T)^3}\equiv \frac{384}{c}\ \frac{\calf}{(\pi T)^3}=-\frac {64}{27}\ \times\
\begin{cases}
1\,,\qquad \langle\calo\rangle=0\Longrightarrow \zet_2\ {\rm is\ unbroken }\,;\\
\kappa(b)\,,\qquad \langle\calo(b)\rangle\ne 0\Longrightarrow \zet_2\ {\rm is\ broken }\,,
\end{cases}
\eqlabel{phases}
\end{equation}
where $\calf$ is the free energy density, $T$ is the temperature and $c$ is the central charge.
The constant $0\le \kappa \le 1$ and the thermal expectation value of $\calo$ (in the symmetry breaking
phase) depends on the parameter $b$ of the dual gravitational
action\footnote{We set the radius of an asymptotic $AdS_4$ 
geometry to unity.}:
\begin{equation}
\begin{split}
S=&\frac{c}{384}\int dx^4\sqrt{-\gamma}\left[R+6
-\frac 12 \left(\nabla\chi\right)^2-2\chi^2-b\chi^4\right]\,,
\end{split}
\eqlabel{spsi}
\end{equation}
The symmetry broken phase exists only $b< b_{crit,0}\equiv -\frac 32$.
Note that since the specific heat density $c_v$ is
\begin{equation}
c_v\equiv -T \left(\frac{\del^2 \calf}{\del T^2}\right)_v=\frac{c\pi^3 T^2}{27}
\ \times\
\begin{cases}
1\,,\qquad \zet_2\ {\rm is\ unbroken }\,;\\
\kappa(b)\ge 0\,,\qquad \zet_2\ {\rm is\ broken }\,,
\end{cases}
\eqlabel{defcv1}
\end{equation}
it is positive, irrespectively whether or not the global symmetry $\zet_2$ is broken.

In this paper we continue study of the model \eqref{spsi}.
First, we point out that there are multiple branches, indexed by $i=0,1,\cdots$,  of the "hair'' --- the thermal
expectation values of $\calo$. In the holographic dual, the index is related to the number of zeros in
the radial profile of the holographic bulk scalar $\chi$. For two branches with $i<j$,
\begin{equation}\
\kappa_i(b)\ >\ \kappa_j(b)\,,\qquad b\in (-\infty,b_{crit,j}<b_{crit,i}]\,, 
\eqlabel{kind}
\end{equation}
\ie the branches with the higher index are increasingly less thermodynamically favored --- all the symmetry
breaking phases have a higher free energy density than that of the $\zet_2$-symmetric phase. 
We find that the thermodynamics of all the symmetry broken phases resemble that of the symmetric phase
in the limit $b\to -\infty$,
\begin{equation}
\biggl(1- \kappa_i(b)\biggr) \propto +\frac{1}{(-b)}\,,\qquad \langle\calo\rangle_i\propto \frac{1}{\sqrt{-b}} \,.
\eqlabel{large}
\end{equation}
Notice that the gravitational potential for $\chi$
\begin{equation}
V\equiv 2\chi^2+b\ \chi^4
\eqlabel{vchi}
\end{equation}
is unbounded from below as $b<0$, (naively) implying that the exotic features of the model
\eqref{spsi} are due to this 'sickness'. This is not the case: given \eqref{large}, it is clear
that a simple deformation of the scalar potential, \ie
\begin{equation}
V\ \to V_g=V+g\ \chi^6\,,\qquad g>0\,,
\eqlabel{vg}
\end{equation}
makes it bounded, while not affecting the thermodynamics of the model, at least  as $b\to -\infty$:
\begin{equation}
V\propto \frac{1}{b}\,,\qquad V-V_g= g\chi^6 \propto \frac{1}{b^3}\,.
\eqlabel{vvg}
\end{equation}
We explicitly demonstrate that the symmetry breaking phases at finite $b$
are robust against the  deformation \eqref{vg}, for small enough $g>0$.

We study the coupled metric-scalar fluctuations in the symmetry broken phases of \eqref{spsi}
and compute the spectrum of the
quasinormal modes (QNMs) of the black branes dual to the conformal order on the lowest
branch\footnote{We expect that the conclusions apply for the $i>0$ branches of the
conformal order as well.}. As expected
from the conformal theory, irrespectively of the symmetry breaking, we find
\begin{equation}
c_s^2=\frac 12\,,\qquad \zeta=0\,,
\eqlabel{cszeta}
\end{equation}
for the speed of the sound waves $c_s$ and the bulk viscosity $\zeta$ correspondingly. 
From the dispersion relation of the
hydrodynamic mode in the shear channel \cite{Kovtun:2005ev} we recover
the universal result \cite{Buchel:2003tz,Kovtun:2004de,Buchel:2004qq} for the ratio of the shear viscosity $\eta$ to the entropy density
$s$ 
\begin{equation}
\frac{\eta}{s}=\frac{1}{4\pi}\,.
\eqlabel{etas}
\end{equation}

Besides the sound wave --- a hydrodynamic QNM in the scalar channel \cite{Kovtun:2005ev} ---
there are two branches of the non-hydrodynamic modes coming from the mixing of the
helicity zero metric fluctuations in the equilibrium black brane geometry and the gravitational
bulk scalar field, whose boundary value determine the order parameter for the $\zet_2$ symmetry
breaking. There are two purely dissipative non-hydrodynamic
modes\footnote{We take the space-time dependence of the QNM fluctuations to
be $\propto e^{-i \omega t +i \vec{k} \vec{x}}$ and introduce $\ww=\frac{\omega}{4\pi T}$ and
$\kk=\frac{|\vec{k}|}{4\pi T}$.}:
\begin{equation}
\Re[\ww] =0\,,\qquad \Im[\ww]\ne 0\,.
\eqlabel{dissnonh}
\end{equation}
One of these QNMs, $\ww_u(\kk)$, at least when the spatial momentum $\kk$ is below some critical
value $\kk_c=\kk_c(b)$, has a positive imaginary part, \ie
\begin{equation}
\Re[\ww_u(\kk)]=0\,,\qquad \Im[\ww_u(\kk)]=
\begin{cases}
\ge 0\,,\qquad \kk\le \kk_c\,;\\
<0\,,\qquad \kk> \kk_c\,.
\end{cases}
\eqlabel{qcrit}
\end{equation}
We further show that
\begin{equation}
\Im[\ww_u(\kk=0)]>0\,,\qquad {\rm as}\qquad b\in (-\infty,b_{crit,0})\,,
\end{equation}
approaching zero in the limit $b\to b_{crit,0}$.
The presence of this mode in the spectrum implies that the translationary invariant horizon
of the black brane dual to a conformal order is perturbatively unstable to clumping.
We expect\footnote{Work in progress.}
that the dynamical evolution of the perturbed conformal order will result in
a destruction of the ordered phase, with the $\zet_2$ symmetric equilibrium phase being
the attractor.  We show that the unstable QNM is present on the higher branches of the
conformal order as well.

Note that the perturbative instability of the holographic conformal order proposed in
\cite{Buchel:2020thm} implies that the Correlated Stability Conjecture of \cite{Gubser:2000ec,Gubser:2000mm}
for the dual black branes is violated: while these black branes have positive specific heat, they are
dynamically unstable\footnote{See \cite{Buchel:2010wk,Buchel:2011ra} for other examples
of the CSC violation.}. We stress that the positive specific heat is not the same as the
thermodynamic stability, tacitly assumed in \cite{Gubser:2000ec} --- the latter concept typically
applies to the thermodynamically dominant phases
(having the minimal free energy density in the canonical ensemble), which is not necessarily the case.
Indeed, unlike the ordered thermodynamically stable phases discovered in \cite{Chai:2020zgq},
the holographic conformal ordered phases in \cite{Buchel:2020thm} are meta-stable, see \eqref{phases}.

A challenge remains to find an example of a {\it stable} holographic thermal conformal order ---
a phase of the black branes which is both the dominant one in the canonical ensemble, and is perturbatively stable with respect
to the linearized fluctuations.

\section{Holographic thermal conformal order}\label{hco}

In this section we review the construction of the holographic conformal order
proposed in \cite{Buchel:2020thm} and discuss two generalizations:
\nxt we show that there are multiple branches of the conformal order;
\nxt we show that the effective scalar potential in the gravitational dual
can be made bounded, without affecting the existence of the ordered phases.

\subsection{Branches of the conformal order}\label{brancheso}

The starting point is the effective action \eqref{spsi}.
The thermal ordered states are dual to black brane solutions
\begin{equation}
ds_4^2=-c_1^2\ dt^2 +c_2^2\ \left[dx_1^2+dx_2^2\right]+ c_3^2\ dr^2\,,
\eqlabel{bbs}
\end{equation}
where all the metric warp factors $c_i$ as well as the bulk scalar $\chi$ are functions of the
radial coordinate $r$,
\begin{equation}
r\ \in\ [r_0,+\infty)\,,
\eqlabel{rrange}
\end{equation}
where $r_0$ is a location of a regular Schwarzschild horizon, and $r\to +\infty$
is the asymptotic $AdS_4$ boundary. Introducing a new radial coordinate
\begin{equation}
x\equiv \frac{r_0}{r}\,,\qquad x\ \in (0,1]\,,
\eqlabel{defx}
\end{equation}
and denoting
\begin{equation}
\begin{split}
&c_1=r \left(1-\frac{r_0^3}{r^3}\right)^{1/2}\ a_1\,,\qquad c_2=r\,,\qquad
c_3=\frac 1r\ \left(1-\frac{r_0^3}{r^3}\right)^{-1/2}\ a_3\,,
\end{split}
\eqlabel{defcs}
\end{equation}
we obtain the following system of ODEs (in a radial coordinate $x$, $'=\frac{d}{dx}$):
\begin{equation}
\begin{split}
&0=a_1'-\frac{3a_1 (a_3^2-1)}{2x (x^3-1)}+\frac18 x a_1 \left(\chi'\right)^2
+\frac{a_3^2 a_1 V}{4x (x^3-1)}\,,
\end{split}
\eqlabel{eq1}
\end{equation}
\begin{equation}
\begin{split}
&0=a_3'+\frac18 x a_3 \left(\chi'\right)^2
+\frac{3a_3 (a_3^2-1)}{2x (x^3-1)}-\frac{a_3^3 V}{4x (x^3-1)}\,,
\end{split}
\eqlabel{eq2}
\end{equation}
\begin{equation}
\begin{split}
&0=\chi''+\left(\frac{a_1'}{a_1}-\frac{a_3'}{a_3}+\frac{x^3+2}{x (x^3-1)}\right)
\chi'+\frac{\del V a_3^2}{(x^3-1) x^2}\,,
\end{split}
\eqlabel{eq3}
\end{equation}
where the scalar potential $V$ is given by \eqref{vchi}, and
$\del V\equiv \frac{\delta V}{\delta \chi}$. 
Notice that $r_0$ is completely scaled out from all the equations of motion.
Eqs.\eqref{eq1}-\eqref{eq3}
have to be solved subject to the following asymptotics:
\nxt in the UV, \ie as $x\to 0_+$,
\begin{equation}
\begin{split}
a_1=1+a_{1,3} x^3+\calo(x^6)\,,\qquad a_3=1-a_{1,3} x^3+\calo(x^6)\,,\qquad \chi=\chi_4 x^4
+\calo(x^7)\,;
\end{split}
\eqlabel{uv}
\end{equation}
\nxt in the IR, \ie as $y\equiv 1-x\to 0_+$,
\begin{equation}
\begin{split}
&a_1=a_{1,0}^h+\calo(y)\,,\qquad a_3=\sqrt{\frac{6}{6-V(c_0^h)}}+\calo(y)\,,\qquad
\chi=c_0^h+\calo(y)\,,
\end{split}
\eqlabel{ir}
\end{equation}
where $V(c_0^h)$ implies that the scalar potential \eqref{vchi} is evaluated on the
horizon value of the bulk scalar $\chi$.

In total, given $b$, the asymptotic expansions are specified by 4 parameters
\begin{equation}
\{a_{1,3}\,,\ \chi_4\,,\ a_{1,0}^h\,,\ c_0^h \}\,,
\eqlabel{allpar}
\end{equation}
which is the correct number of parameters necessary to provide a solution to a system
of a single second order and two first order equations, $1\times 2 +2 \times 1=4$.
It is straightforward to extract the thermodynamics of the resulting black brane:
\begin{equation}
\begin{split}
&\calf=-\frac{c}{192}\ (\pi T)^3\ \kappa\,,\qquad \cale=-2\calf\,,\qquad \frac{\langle\calo\rangle}{c T^4}
\propto \langle\hat{\calo}\rangle=\chi_4\,,\\
&T=\frac{r_0}{8\pi}\ a_{1,0}^h\ \sqrt{36-6 V(c_0^h)}\,,\qquad
\kappa=\frac{36}{(a_{1,0}^h)^2(36-6 V(c_0^h))}\,,
\end{split}
\eqlabel{vevs}
\end{equation}
for the free energy density $\calf$, the energy density $\cale$, the temperature $T$, and the thermal expectation value
of the conformal order $\langle\hat{\calo}\rangle$ . We explicitly indicated
the $r_0$ dependence --- all the parameters in \eqref{allpar}, as well as $\kappa$ and
$V(c_0^h)$ depend only on $b$. 

From \eqref{vevs}, the speed of the sound waves in the holographic conformal order plasma
is
\begin{equation}
c_s^2=-\frac{\del \calf}{\del \cale}=\frac 12\,.
\eqlabel{cs2}
\end{equation}
Additionally, since that temperature $T$ depends on $r_0$, and the horizon value of the
scalar is $r_0$-independent,
\begin{equation}
\frac{d}{dT} \biggl(\chi(x)\bigg|_{x\to 1_-}\biggr)=\frac{d}{dT} \biggl(c_0^h\biggr)=0\,,
\end{equation}
implying that the bulk viscosity of the corresponding plasma must
vanish\footnote{Eling-Oz formula implies that the holographic plasma bulk viscosity
is proportional to the square of the derivative of the bulk scalar field(s) evaluated at the
horizon  with respect to the temperature, keeping all the mass parameters fixed
\cite{Buchel:2011yv,Buchel:2011wx}. 
} \cite{Eling:2011ms}
\begin{equation}
\frac{\zeta}{T^2}=0\,.
\eqlabel{zeta}
\end{equation}
In section \ref{transport} we reproduce \eqref{cs2} and \eqref{zeta} from the
dispersion relation of the hydrodynamic QNMs. 

Note that the "disordered phase'' corresponds to identical vanishing of the bulk scalar
field, in this case
\begin{equation}
\chi_4=c_0^h=0\,,\ a_{1,0}^h=1 \qquad \Longrightarrow\qquad  \kappa_{disordered}=1\,,\
\langle\hat{\calo}\rangle=0 \,. 
\eqlabel{vanilla}
\end{equation}
To understand the ordered phases it is the easiest to consider $b\to -\infty$ limit.
From \eqref{eq1}-\eqref{eq3} it is straightforward to see that
\begin{equation}
\chi(x)=\sum_{n=1}^\infty\ \frac{f_{[2n-1]}(x)}{(-b)^{n-1/2}}\,,\qquad
a_1(x)=1+\sum_{n=1}^\infty \frac{a_{1,[n]}(x)}{(-b)^n}\,,\qquad
a_3(x)=1+\sum_{n=1}^\infty \frac{a_{3,[n]}(x)}{(-b)^n}\,,
\eqlabel{perlimit}
\end{equation}
and correspondingly (from \eqref{uv} $a_{1,3}\equiv -a_{3,3}$)
\begin{equation}
\begin{split}
&a_{3,3}=\sum_{n=1}^\infty \frac{a_{3,[n],3}}{(-b)^n}\,,\
\chi_4=\sum_{n=1}^\infty\ \frac{f_{[2n-1],4}}{(-b)^{n-1/2}}\,,\
c_0^h=\sum_{n=1}^\infty\ \frac{f_{[2n-1],0}^h}{(-b)^{n-1/2}}\,,\ a_{1,0}^h=1+\sum_{n=1}^\infty\frac{a_{1,[n],0}^h}{(-b)^n}\,,
\end{split}
\eqlabel{perlimit2}
\end{equation}
\ie in this limit the ``hairy'' black branes approach the standard $AdS_4$-Schwarzschild
solution\footnote{As we will see in section \ref{instability},
this is not the case for the spectrum of the non-hydrodynamic QNMs: some
of the QNMs of the hairy black branes remain distinct from the $AdS_4$-Schwarzschild
black brane QNMs in the limit $b\to -\infty$.}. To leading order, \ie $n=1$, we
have:
\begin{equation}
\begin{split}
&0=f_{[1]}''+\frac{x^3+2}{x (x^3-1)} f_{[1]}'-\frac{4 (f_{[1]}^2-1) f_{[1]}}{x^2 (x^3-1)}\,,
\end{split}
\eqlabel{peq1}
\end{equation}
\begin{equation}
\begin{split}
&0=a_{3,[1]}'+\frac{3 a_{3,[1]}}{x (x^3-1)}+\frac{x}{8} (f_{[1]}')^2
+\frac{f_{[1]}^2 (f_{[1]}^2-2)}{4x (x^3-1)}\,,
\end{split}
\eqlabel{peq2}
\end{equation}
\begin{equation}
\begin{split}
&0=a_{1,[1]}'-\frac{3 a_{3,[1]}}{x (x^3-1)}+\frac x8 (f_{[1]}')^2
-\frac{f_{[1]}^2 (f_{[1]}^2-2)}{4x (x^3-1)}\,.
\end{split}
\eqlabel{peq3}
\end{equation}

\begin{figure}[t]
\begin{center}
\psfrag{k}[cc][][1.0][0]{$\kappa$}
\psfrag{f}[cc][][0.7][0]{$\lim_{b\to -\infty}({-b})^{1/2}\ \chi(x)$}
\psfrag{x}[cc][][1.0][0]{$x$}
\psfrag{b}[cc][][1.0][0]{$b$}
\includegraphics[width=3in]{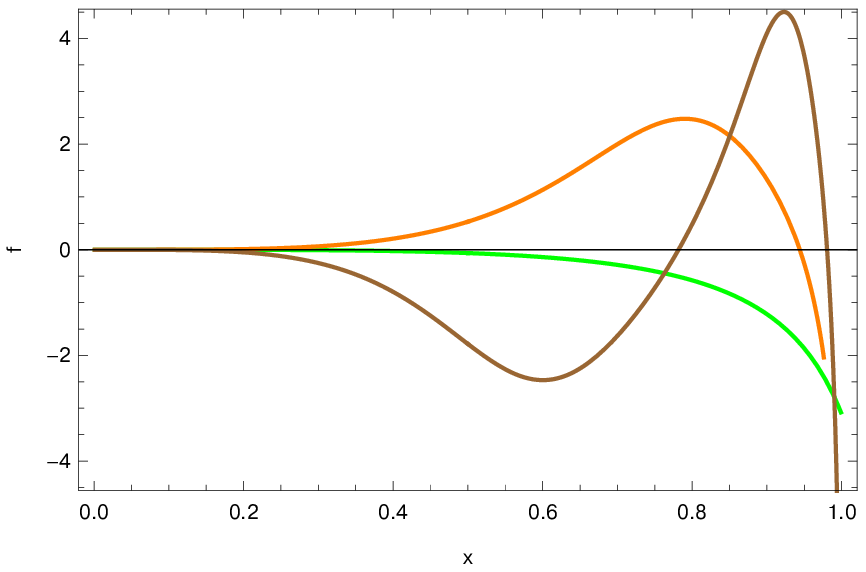}\,
\includegraphics[width=3in]{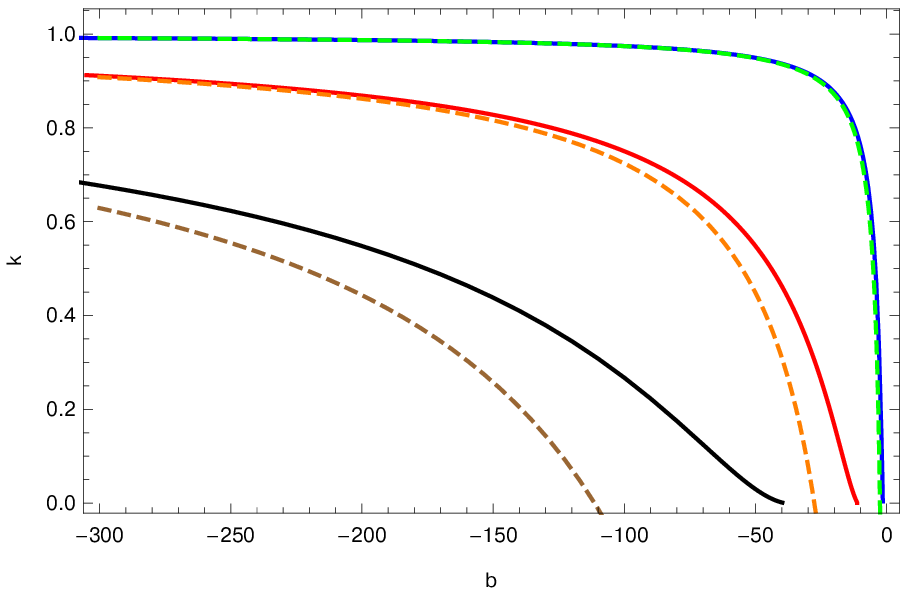}
\end{center}
  \caption{Left panel: branches (``overtones'') of the holographic conformal
  order are characterized by the number of zeros of the gravitational scalar dual
  to the order parameter. We plot the scalar profiles as $b\to -\infty$.
  The green/orange/brown profiles correspond to an index $i=0,1,2$ branch/overtone.
  Right panel: $\kappa(b)$ characterizing the thermodynamics of the
  ordered phases (see eq.~\eqref{phases}) for $i=0,1,2$ (solid blue/red/black curves correspondingly).
  The dashed green/orange/brown curves represent the leading order $b\to -\infty$
  approximation, see eq.~\eqref{lkappa}.
} \label{fig1}
\end{figure}

There is a discrete set of solutions of \eqref{peq1} subject to the boundary conditions inherited from
\eqref{uv} and \eqref{ir}, characterized by the number of zeroes in the function $f_{[1]}$. In the left panel of
fig.~\ref{fig1} we present
\begin{equation}
f_{[1]}(x)\ \equiv\ \lim_{b\to -\infty} (-b)^{1/2} \chi(x) \,,
\eqlabel{plf1}
\end{equation}
with $i=0,1,2$ (green, orange, brown curves correspondingly) zeros. The discrete set of solutions for
\eqref{peq1} leads to discrete sets of $a_{3,[1]}$ and $a_{1,[1]}$. We just constructed the
lowest\footnote{Only this overtone was discussed in \cite{Buchel:2020thm}. } ($i=0$) and the
higher $(i=1,2)$ overtones of the holographic conformal order:
\begin{center}
\[
\begin{tabular}{|c|c|c|c|c|}
\hline
$\#\ {\rm of\ zeros}$  &  $f_{1,[4]}$ & $f_{[1],0}^h$ & $a_{3,[1],3}$ & $a_{1,[1],0}^h$ \\
\hline\hline
$0$  &  $\pm 0.914 $ &  $\pm 3.114$ &  $ 0.644 $ &  $-4.928 $ \\
\hline
$1$  &  $\pm 7.875$ &   $\mp 6.789$ &  $6.899$  &  $ -155.534$ \\
\hline
$2$  &  $\pm 30.546$ & $\pm 11.233$ &  $27.835$ & $-1249.88$\\
\hline
\end{tabular}
\]
\[
\]
\end{center}
where $\pm$ for the parameters specifying $f_{[1]}$ reflects the spontaneously broken global $\zet_2$ symmetry.  
To leading order as $b\to -\infty$ (see \eqref{vevs})
\begin{equation}
\kappa=1+\frac{2 a_{1,[1],0]}^h+\frac16 (f_{[1],0}^h)^4-\frac13 (f_{[1],0}^h)^2)}{b}+\calo\left(\frac{1}{b^2}\right)\,,
\eqlabel{lkappa}
\end{equation}
which is represented by the dashed green/orange/brown (for $i=0,1,2$ correspondingly)
curves in the right panel of fig.~\ref{fig1}. Once the overtones of the conformal order are constructed
in the limit $b\to -\infty$, it is straightforward to solve eqs.~\eqref{eq1}-\eqref{eq3} and extend the results
for $\kappa_i$ to finite values of $b$. This is shown with the solid blue/red/black
(for $i=0,1,2$ correspondingly) curves in the right panel. For each overtone of the conformal order there is
a critical value of $b$, \ie $b_{crit,i}$, beyond which the overtone disappears from the
spectrum\footnote{In all cases as $b\to b_{crit,i}$ the order parameter $\langle\hat{\calo}\rangle_i$
diverges, see also \cite{Buchel:2020thm}.}:
\begin{center}
\[
\begin{tabular}{|c||c|c|c|}
\hline
$i=\#\ {\rm of\ zeros}$ &   $0$ & $1$ & $2$  \\
\hline
$b_{crit,i}$  &  $-1.5$   &$-11.258$ &  $-39.295$\\
\hline
\end{tabular}
\]
\[
\]
\end{center}
Note that
\begin{equation}
\lim_{b\to b_{crit,i}} \kappa_i(b) =0\,,
\eqlabel{klimit}
\end{equation}
and for each $i<j$,
\begin{equation}
1>\kappa_i(b)\ > \kappa_j(b)\qquad {\rm and}\qquad  b_{crit,i}<b_{crit,j} \,.
\eqlabel{klimit2}
\end{equation}
Thus, all the ordered phases are subdominant (have the higher free energy density) compare to the
$\zet_2$-symmetric phase, see \eqref{vanilla}. Additionally, the free energy density of the conformal
order overtones increases (at fixed $b$) with its index.  

Equilibrium thermal phases with or without the global $\zet_2$ symmetry have positive specific heat.
There is a latent heat $\Delta \cale$ associated with the transitions between the symmetry broken phases $i<j$,
and the transitions to the $\zet_2$ symmetric phase,
\begin{equation}
\Delta\cale_{j\to i}\ \propto +(\kappa_i-\kappa_j) T^3\,,\qquad \Delta\cale_{i\to disordered}\ \propto +(1-\kappa_i) T^3\,,
\eqlabel{latent}
\end{equation}
typically indicative of the first-order phase transition. Rather, as we show in section \ref{instability},
each of the ordered phases suffers the perturbative instability for any value of $b$ it exists.
A natural guess is that the end point of the dynamical evolution will bring us from the ordered phase to a
disordered, $\zet_2$-symmetric, phases. However, to confirm this, a numerical simulation
is necessary\footnote{Another possibility is the evolution
to a naked singularity in a dual gravitational description, see \cite{Bosch:2017ccw,Buchel:2017map}.}.

\subsection{Holographic conformal order with the bounded gravitational potential}

\begin{figure}[t]
\begin{center}
\psfrag{k}[cc][][1.0][0]{$\kappa$}
\psfrag{c}[cc][][1.0][0]{$|\langle \hat\calo\rangle|^{-1}$}
\psfrag{g}[cc][][1.0][0]{$g$}
\includegraphics[width=3in]{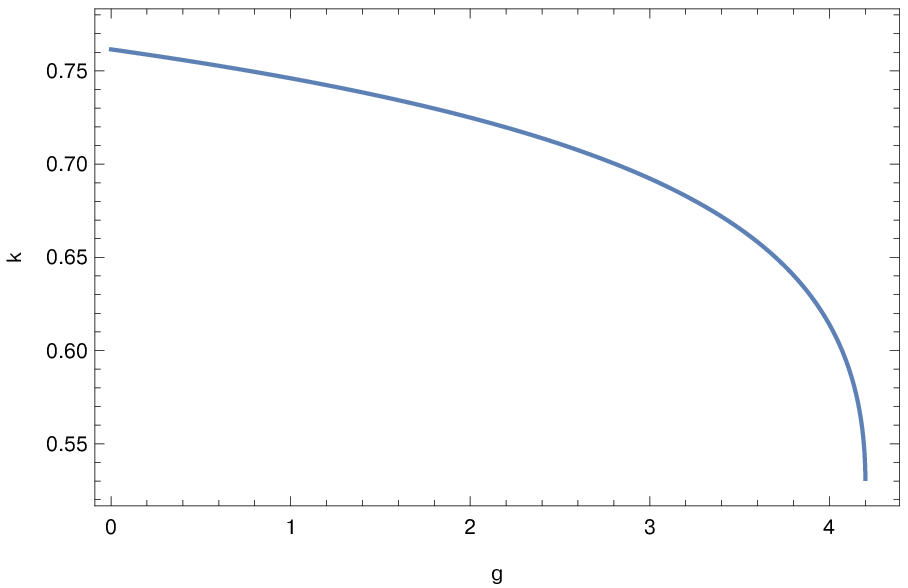}\,
\includegraphics[width=3in]{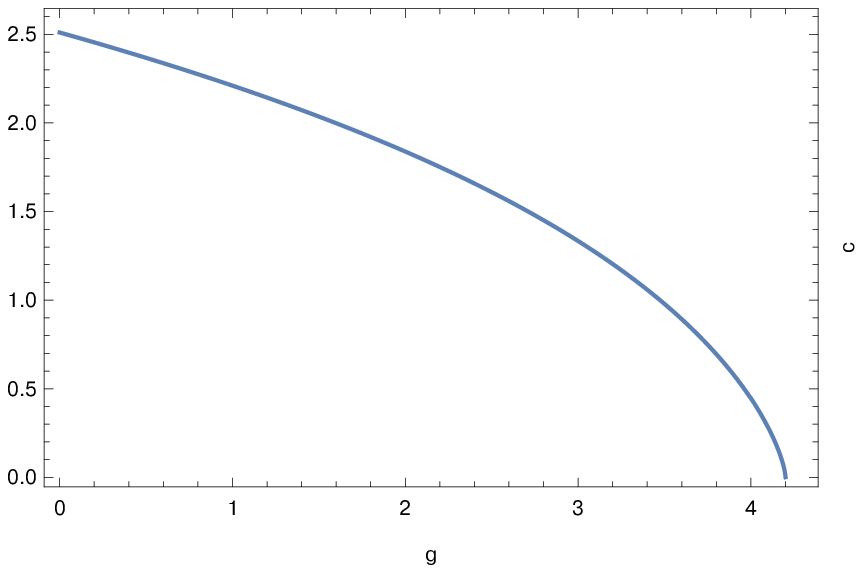}
\end{center}
  \caption{Left panel: $\kappa$ parameter of the lowest branch of the conformal order
  at $b=-10$ as a function of $g$, a deformation parameter bounding the potential
  of the bulk scalar dual to the order parameter, see eq.~\eqref{vg}. Holographic
  conformal order persists for $0\le g<g_{crit} \approx 4.2$. Right panel: the corresponding
  dependence of  the thermal order parameter $\langle\hat\calo\rangle$. Note that
  the order parameter diverges as $g\to g_{crtit}$.
} \label{fig2}
\end{figure}

The gravitational potential for a scalar field dual to a conformal order parameter
is unbounded from below, see \eqref{vchi} and note that the existence of the conformal
order requires $b< -\frac 32$. We used  simple scaling arguments in section \ref{intro}
to suggest that the conformal order exists once the scalar potential is made bounded as in
\eqref{vg}, at least for sufficiently small $g$ and in the limit $b\to -\infty$.

What happens at finite $b$? First, note that all the analysis in section \eqref{hco}
allow for a simple generalization as $V$ is replaced with $V_g$ of \eqref{vg} ---
see eqs.~\eqref{eq1}-\eqref{eq3} and \eqref{vevs}. In the left panel of fig.~\ref{fig2}
we show the results for the parameter $\kappa$ characterizing the thermodynamics of the
lowest branch of the
conformal order (see \eqref{phases}) evaluated at $b=-10$ as a function of $g>0$.
We find that the holographic conformal order persists for
\begin{equation}
g\ \in\ [0,g_{crit})\,,\qquad g_{crit}\bigg|_{b=-10}\approx 4.2\,.
\eqlabel{grange}
\end{equation}
As $g\to g_{crit}$, the order parameter $|\langle\hat\calo\rangle|$ diverges, see the
right panel of fig.~\ref{fig2}.

It is clear that the conformal order constructed in section \ref{hco} is robust
against the deformation of the bulk scalar potential as in \eqref{vg} for $0\le g< g_{crit}(b)$
since the basic equations determining it, \ie eqs.~\eqref{eq1}-\eqref{eq3} and \eqref{vevs},
are analytic in the limit $g\to 0_+$.

\section{QNMs of the conformal order}

In this section we study the spectrum of the linearized fluctuations about hairy black branes,
dual to the conformal order --- the spectrum of the quasinormal modes \cite{Kovtun:2005ev}:
\nxt In section \ref{transport} we consider the hydrodynamic modes, \ie QNMs such that
$\ww(\kk)\to$ as $\kk\to 0$. We confirm the conformal speed of the sound waves \eqref{cs2},
the vanishing of the bulk viscosity \eqref{zeta}, and the universality of the ratio of the shear
viscosity to the entropy density \cite{Buchel:2003tz,Kovtun:2004de,Buchel:2004qq}.
\nxt In section \ref{instability} we consider the spectrum of the non-hydrodynamic modes.
We discuss different branches of the spectra and exhibit the QNM with $\Im [\ww]>0$,
for $\kk\le \kk_c$, see eq.~\eqref{qcrit}. This mode make holographic conformal order
of \cite{Buchel:2020thm} perturbatively unstable. Its existence is yet another
counterexample of the correlated stability conjecture \cite{Gubser:2000ec,Gubser:2000mm}.

For the most part we focus on the QNM spectra of the lowest branch of the conformal order
with a dual gravitational action \eqref{spsi}. We discuss however the instability of the
higher overtones of the conformal order.

Consider fluctuations of the background geometry \eqref{bbs}
\begin{equation}
g_{\mu\nu} \to g_{\mu\nu}+h_{\mu\nu}\,,\qquad \chi\to \chi+f\,.
\eqlabel{gflu}
\end{equation}
For convenience, we partially fix the gauge by requiring
\begin{equation}
h_{tr}=h_{x_i r}=h_{rr}=0\,.
\eqlabel{gaugefix}
\end{equation}
We orient the coordinate system in such a way that the $x_2$ axis is directed along the spatial
momentum, and assume that all the fluctuations depend only on $(t,x_2,r)$, \ie we have
a $\zet_2$ parity symmetry along the $x_1$-axis. At a linearized level, the following  
sets of fluctuations decouple from each other
\begin{equation}
\begin{split}
\zet_2-{\rm even}:\qquad\qquad\qquad&  \{h_{tt},h_{tx_2},h_{x_1x_1},h_{x_2x_2}; f\}\,;\\
\zet_2-{\rm odd}:\qquad\qquad\qquad&  \{h_{tx_1},h_{x_1x_2}\}\,.
\end{split}
\eqlabel{decoupling}
\end{equation}
Let 
\begin{equation}
\begin{split}
&h_{tt}=-e^{-i \omega t+i k x_2}\ c_1(r)^2\ H_{tt}\,, \qquad h_{x_ix_i}=e^{-i \omega t+i k x_2}\ c_2(r)^2\ H_{x_ix_i}\,,\\
&h_{tx_i}=e^{-i \omega t+i k x_2}\ c_2(r)^2\ H_{tx_i}\,,\qquad h_{x_1x_2}=e^{-i \omega t+i k x_2}\ c_2(r)^2\ H_{x_1x_2}\,,\\
&f=e^{-i \omega t+i k x_2}\ F\,,
\end{split}
\eqlabel{flrescale}
\end{equation}
where $\{H_{tt},H_{tx_i},H_{x_ix_i},H_{x_1x_2},F\}$ are functions of the radial coordinate only and
$c_i(r)$ are defined in \eqref{defcs}. Following \cite{Kovtun:2005ev,Benincasa:2005iv},
we introduce fluctuations invariant under the residual diffeomorphisms preserving
\eqref{gaugefix}:
\begin{itemize}
\item $\zet_2$-even, the {\it sound channel},
\begin{equation}
\begin{split}
&Z_H\equiv 4 \frac k\omega\ H_{tx_2}+2 H_{x_2x_2}-2 H_{x_1x_1}
\left(1-\frac{k^2}{\omega^2}\  \frac{(c_1^2)'}{(c_2^2)'}\right)-2 \frac{k^2}{\omega^2}\
\frac{c_1^2}{c_2^2}\ H_{tt}\,,\\
&Z_F\equiv f-\frac{\chi'}{(\ln c_2^2)'}\ H_{x_1x_1}\,;
\end{split}
\eqlabel{gaugesound}
\end{equation}
\item $\zet_2$-odd, the {\it shear channel},
\begin{equation}
\begin{split}
&Z_s\equiv k H_{tx_1}+\omega H_{x_1x_2}\,.
\end{split}
\eqlabel{gaugeshear}
\end{equation}
\end{itemize}

Using the radial coordinate \eqref{defx}, we find the following equations of motion:
\nxt for the sound channel,
\begin{equation}
\begin{split}
&0=Z_H''+\cala_H\ Z_H'+\calb_H\ Z_H+\calc_H\ Z_F\,,
\end{split}
\eqlabel{s1}
\end{equation}
\begin{equation}
\begin{split}
&0=Z_F''+\cala_F\ Z_F'+\calb_F\ Z_H'+\calc_F\ Z_F+\cald_F\ Z_H\,,
\end{split}
\eqlabel{s2}
\end{equation}
with
\begin{equation}
\begin{split}
&\cala_H=\biggl(
a_1^2 \tk^2 x^4 (x^3-1)^2 (\chi')^4+a_1^2 \tk^2 x^2 (x^3-1)
(a_3^2 \tV+6 (1-x^3)) (\chi')^2
\\&-2 a_3^4 a_1^2 \tk^2 \tV^2
+8 a_3^2 (a_1^2 \tk^2 (x^3-1)-2 \tw^2) \tV-8 (x^3-1) (5 a_1^2 \tk^2 (x^3-1)\\
&-4 \tw^2)
\biggr)
\biggl(2 x (x^3-1)  (a_1^2 \tk^2 x^2 (x^3-1) (\chi')^2+2 a_3^2 a_1^2 \tk^2 \tV
+4 (a_1^2 \tk^2 (x^3-1)\\
&+4 \tw^2))
\biggr)^{-1}\,,
\end{split}
\eqlabel{ah}
\end{equation}
\begin{equation}
\begin{split}
&\calb_H=\tw^2 \biggl(
a_1^4 \tk^2 x^6 (x^3-1)^3 (\chi')^6+4 a_1^4 \tk^2 x^4 (x^3-1)^2
(a_3^2 \tV+5 (1-x^3)) (\chi')^4\\
&+4 a_1^2 \tk^2 x^2 (x^3-1)
(a_3^4 a_1^2 \tV^2-8 a_3^2 a_1^2 (x^3-1) \tV+4 a_3^2 \tw^2 x^2
+4 a_1^2 (x^3-1) (a_3^2 \tk^2\\
&\times x^2+3 (x^3-1)))
(\chi')^2+16 a_3^4 a_1^4 \tk^2 (x^3-1) \tV^2+32 a_3^2 a_1^2 \tk^2 (a_1^2 (x^3-1) (a_3^2 \tk^2 x^2
\\&+6 (1-x^3))+a_3^2 \tw^2 x^2) \tV+64 a_1^4 \tk^2 (x^3-1)^2 (a_3^2 \tk^2 x^2+9 (x^3-1))\\
&+320 a_1^2 \tk^2 x^2 a_3^2 (x^3-1) \tw^2+256 a_3^2 \tw^4 x^2
\biggr) \biggl(16 a_1^2 (x^3-1)^2 (a_1^2 \tk^2 \tw^2 x^4 \\
&\times (x^3-1) (\chi')^2+2 x^2 \tw^2 \tk^2 a_1^2 a_3^2 \tV+4 x^2 \tw^2 (a_1^2 \tk^2 (x^3-1)+4 \tw^2))
\biggr)^{-1}\,,
\end{split}
\eqlabel{bh}
\end{equation}
\begin{equation}
\begin{split}
&\calc_H=a_1^2 \tk^2 \biggl(
(\chi')^2 x^2 (x^3-1)+2 a_3^2 \tV+12 (1-x^3)\biggr)
\biggl(
\tw^2 x^3 (x^3-1) (\chi')^3\\&+\frac14 a_1^2 \tk^2 x^2 a_3^2(x^3-1) \tdV (\chi')^2-2 (x^3-1)
(6 x \tw^2+a_3^2 a_1^2 \tk^2 x \tV) \chi'+a_3^2 (a_1^2 \tk^2 \\&\times (x^3-1)+4 \tw^2) \tdV
+\frac12 a_3^4 a_1^2 \tk^2 \tV \tdV
\biggr)
\biggl(
(x^3-1) x^2 \tw^2 (a_1^2 \tk^2 x^2 (x^3-1)\\&\times (\chi')^2+2 a_3^2 a_1^2 \tk^2 \tV
+4 a_1^2 \tk^2 (x^3-1)+16 \tw^2)
\biggr)^{-1}\,,
\end{split}
\eqlabel{ch}
\end{equation}
\begin{equation}
\begin{split}
&\cala_F=-\frac{a_3^2 \tV+2 (1-x^3)}{2(x^3-1) x}\,,
\end{split}
\eqlabel{af}
\end{equation}
\begin{equation}
\begin{split}
&\calb_F=2 a_3^2 \tw^2 \biggl(\chi' x \tV-2 \tdV\biggr)
\biggl((x^3-1) x (a_1^2 \tk^2 x^2 (x^3-1) (\chi')^2+2 a_3^2 a_1^2 \tk^2 \tV\\
&+4 a_1^2 \tk^2 (x^3-1)+16 \tw^2)
\biggr)^{-1} \,,
\end{split}
\eqlabel{bf}
\end{equation}
\begin{equation}
\begin{split}
&\calc_F=-a_3^2 \biggl(
\frac12 a_1^4 \tk^2 x^3 (x^3-1)^2 \tdV (\chi')^3
+(-4 a_1^2 x^2 (x^3-1) (a_1^2 \tk^2 (x^3-1)+\tw^2)\\& \times \tV
-a_1^4 \tk^2 x^2 (x^3-1)^2 \del^2\tV-a_1^2 \tk^2 x^2 (x^3-1)
(a_1^2 (x^3-1) (\tk^2 x^2+4)\\&+\tw^2 x^2)) (\chi')^2+a_1^2 x 
(x^3-1) \tdV (a_3^2 a_1^2 \tk^2 \tV+10 a_1^2 \tk^2 (x^3-1)+16 \tw^2) \chi'
\\&-2 (a_1^2 (x^3-1) \del^2\tV
+a_1^2 (x^3-1) (\tk^2 x^2+4)+\tw^2 x^2) (a_3^2 a_1^2 \tk^2 \tV+2 a_1^2 \tk^2\\
&\times(x^3-1)
+8 \tw^2)
\biggr)
\biggl(
(x^3-1)^2 a_1^2 x^2 
(a_1^2 \tk^2 x^2 (x^3-1) (\chi')^2+2 a_3^2 a_1^2 \tk^2 \tV+4 a_1^2\\&\times \tk^2 (x^3-1)+16 \tw^2)
\biggr)^{-1}\,,
\end{split}
\eqlabel{cf}
\end{equation}
\begin{equation}
\begin{split}
&\cald_F=a_3^2 \tw^2 \biggl(x \tV \chi'-2 \tdV\biggr)
\biggl(
(\chi')^2 x^2 (x^3-1)+2 a_3^2 \tV+12 (1-x^3)\biggr)
\biggl(4 (x^3-1)^2 x^2\\&\times (a_1^2 \tk^2 x^2 (x^3-1) (\chi')^2+2 a_3^2 a_1^2 \tk^2 \tV
+4 a_1^2 \tk^2 (x^3-1)+16 \tw^2)
\bigg)^{-1}\,,
\end{split}
\eqlabel{df}
\end{equation}
where we introduced
\begin{equation}
\begin{split}
&\tV=V-6\,,\qquad \tdV=\frac{\delta \tV}{\delta \chi}\,,\qquad \del^2\tV=\frac{\delta^2 \tV}
{\delta \chi^2}\,,\\
&\kk=\frac{k}{4\pi T}=\frac{k}{r_0}\ \frac{1}{\alpha}\,,\qquad \ww=\frac{\omega}{4\pi T}=
\frac{\omega}{r_0}\ \frac{1}{\alpha}\,,\qquad \alpha=a_{1,0}^h\sqrt{9-\frac 32 V(c_0^h)}\,,
\end{split}
\eqlabel{tildedef}
\end{equation}
with $V$ being the scalar potential \eqref{vchi};
\nxt and for the shear channel,
\begin{equation}
\begin{split}
&0=Z_s''+\cala_s\ Z_s'+\calb_s\ Z_s \,,
\end{split}
\eqlabel{ss}
\end{equation}
with
\begin{equation}
\begin{split}
&\cala_s=\biggl(
a_1^2 \tk^2 x^2 (x^3-1)^2 (\chi')^2-2 a_3^2 \tw^2 \tV-(4 (x^3-1)) (2 a_1^2 \tk^2 (x^3-1)\\
&-\tw^2)\biggr)\biggl(
4 x (x^3-1) (a_1^2 \tk^2 (x^3-1)+\tw^2)\biggr)^{-1}\,,
\end{split}
\eqlabel{as}
\end{equation}
\begin{equation}
\begin{split}
&\calb_s=\frac{a_3^2 (a_1^2 \tk^2 (x^3-1)+\tw^2)}{(x^3-1)^2 a_1^2}\,.
\end{split}
\eqlabel{bs}
\end{equation}

\subsection{Hydrodynamic modes and the  transport}\label{transport}

We begin with the $\kk\to 0$ limit of the dispersion relation for the
shear and the sound channels QNM modes. From the former, we extract the shear viscosity, and
from the latter, the speed of the sound waves and the bulk viscosity of the
holographic conformal order. We discuss the dispersion relation
$\ww=\ww(\kk)$ for the sound waves in section \ref{dis} for different
values of $b$.

\subsubsection{The shear viscosity}

The shear mode dispersion relation in the limit $\kk\to 0$ takes form
\begin{equation}
\ww=-i\ \frac{4\pi\eta}{s}\ \kk^2+\calo(\kk^4)\,.
\eqlabel{sheardis}
\end{equation}
We now evaluate \eqref{sheardis} in the holographic conformal order. To this end,
we set
\begin{equation}
Z_s=x^3 (1-x)^{-i \ww}\left(Z_{s,0}+\kk^2 Z_{s,2}+\calo(\kk^4)\right)\,,\qquad \ww=-i\ \beta\ \kk^2
+\calo(\kk^4)\,.
\eqlabel{shearq}
\end{equation}
Using \eqref{ss}, we find
\begin{equation}
\begin{split}
&0=Z_{s,0}''+\frac{x^2 (\chi')^2+16}{4x}\ Z_{s,0}'+\frac34 (\chi')^2\ Z_{s,0}\,, \\
&0=Z_{s,2}''+\frac{x^2 (\chi')^2+16}{4x}\ Z_{s,2}'+\frac34 (\chi')^2\ Z_{s,2}+J_{s,2}\,,\\
\end{split}
\eqlabel{sheareqs}
\end{equation}
where
\begin{equation}
\begin{split}
&J_{s,2}=\beta \biggl(
\frac{x \beta(\chi')^2}{4a_1^2 (x^3-1)} +\frac{\beta a_3^2 \tV}{2x (x^3-1)^2 a_1^2}
-\frac{3 \beta}{a_1^2 (x^3-1) x}-\frac{2}{x-1}\biggr)\ Z_{s,0}'-\biggl( (\chi')^2\\
&\times
\frac{\beta (a_1^2 x (x^2+x+1)-3 \beta)}{4(x^3-1) a_1^2}-\frac{a_3^2 \alpha^2}
{x^3-1}+\frac{(3 x-4) \beta}{x (x-1)^2}-\frac{(3 a_3^2 \tV-18 (x^3-1)) \beta^2}
{2 (x^3-1)^2 a_1^2 x^2}
\biggr)\ Z_{s,0}\,.
\end{split}
\eqlabel{sheareqs1}
\end{equation}
Equations \eqref{sheareqs} have to be solved subject to the
following boundary conditions:
\nxt in the UV, \ie as $x\to 0_+$, 
\begin{equation}
Z_{s,0}=1+\calo(x^8)\,,\qquad Z_{s,2}=-\beta x+\calo(x^2)\,;
\eqlabel{shearbc1}
\end{equation}
\nxt in the IR, \ie as $y\equiv 1-x\to 0_+$, 
\begin{equation}
Z_{s,0}=z_{0,0}^h+z_{0,1}^h\ y+\calo(y^2)\,,\qquad Z_{s,2}=z_{2,0}^h+z_{2,1}^h\ y+\calo(y^2)\,.
\eqlabel{shearbc2}
\end{equation}
Given \eqref{shearq}, these asymptotes reflects the incoming boundary conditions
as the horizon, and the Dirichlet boundary conditions at the asymptotic boundary
\cite{Kovtun:2005ev}.

\begin{figure}[t]
\begin{center}
\psfrag{b}[cc][][1.0][0]{$b$}
\psfrag{c}[cc][][1.0][0]{${4\pi\ \eta/ s}-1$}
\includegraphics[width=5in]{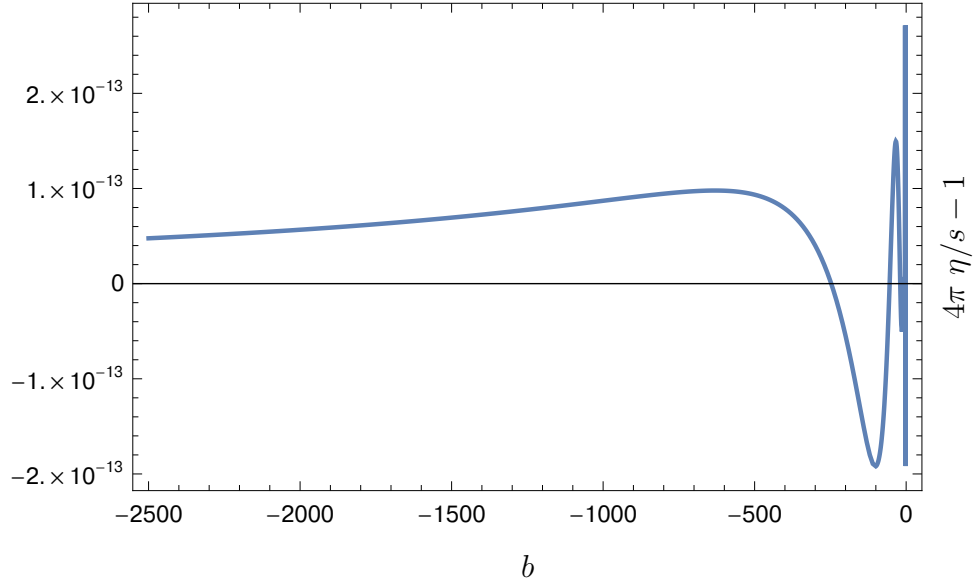}
\end{center}
  \caption{The ratio of the shear viscosity to the entropy density in the holographic
  conformal order as a function of $b$ obtained from the analysis of the
  shear channel QNMs. 
} \label{fig3}
\end{figure}

Ultimately, we need to extract $\beta=\beta(b)$. First, we need to solve numerically the
first equation in \eqref{sheareqs}, producing the data sets $\{z_{0,0}^h(b),z_{0,1}^h(b)\}$.
Remarkably, we do not need to solve for $Z_{s,2}$: direct series expansion of
the second equation in \eqref{sheareqs} in the IR leads to
\begin{equation}
\begin{split}
&0=\frac{(3 (a_{1,0}^h)^2 z_{0,0}^h-3 \beta z_{0,0}^h+\beta z_{0,1}^h) \beta}
{3(a_{1,0}^h)^2}\  \frac{1}{y^2}+\frac 1y\  \frac{3 (a_{1,0}^h)^2 z_{0,0}^h
-3 \beta z_{0,0}^h+\beta z_{0,1}^h}{3(c_0^h)^2 \tV(c_0^h)^2 (a_{1,0}^h)^2 } \biggl(
16 \beta ((c_0^h)^2\\&-6)
((c_0^h)^2-6-2 \tV(c_0^h))-(3 (a_{1,0}^h)^2 (c_0^h)^2-\beta (c_0^h)^2-16 \beta)
\tV(c_0^h)^2\biggr) +\calo(y^0)\,,
\end{split}
\eqlabel{yexp}
\end{equation}
with the parameters of the asymptotic expansion of $Z_{s,2}$ (see \eqref{shearbc2})
entering only in $\calo(y)$. Thus, we find  
\begin{equation}
\beta=\frac{3 (a_{1,0}^h)^2 z_{0,0}^h}{3 z_{0,0}^h- z_{0,1}^h}\,.
\eqlabel{betares}
\end{equation}
In the disordered phase, see eq.~\eqref{vanilla}, the  equation for $Z_{s,0}$ is very simple:
\begin{equation}
\chi(x)\bigg|_{disordered}\equiv 0\qquad \Longrightarrow\qquad
Z_{s,0}\bigg|_{disordered}\equiv 1 \qquad \Longrightarrow\qquad \beta|_{disordered}=1\,,
\eqlabel{bdis}
\end{equation}
leading to the universal result for the ratio of the shear viscosity to the entropy density
\cite{Buchel:2003tz}.  In the symmetry broken phase, from the
perspective of the shear channel QNM discussed here,
we do not have a general proof why $\beta$ must be
unity as well. In fig.~\ref{fig3} we present $\beta-1$ of \eqref{betares}, from the
numerical solution of the equation for $Z_{s,0}$.

We would like to stress that we already know from \cite{Buchel:2003tz}
that\footnote{The universality of the shear viscosity in the holographic plasma was never
proven from the perspective of the shear channel QNMs dispersion relation.} $\beta=1$;
the analysis presented here should be viewed as highly nontrivial check on the QNM equations
and our numerical construction of the holographic order parameter.

\subsubsection{The speed of the sound waves and the bulk viscosity}

The sound mode dispersion relation in the limit $\kk\to 0$ takes form
\begin{equation}
\ww=c_s\ \kk -\frac i2\ \frac{4\pi\eta}{s}\ \biggl(1+\frac{\zeta}{\eta}\biggr)\ \kk^2+\calo(\kk^3)\,.
\eqlabel{sounddis}
\end{equation}
We now evaluate \eqref{sounddis} in the holographic conformal order. To this end,
we set
\begin{equation}
\begin{split}
&Z_H=(1-x)^{-i \ww}\left(Z_{H,0}+i\ \kk Z_{H,1}+\calo(\kk^2)\right)\,,\\
&Z_F=(1-x)^{-i \ww}\left(Z_{F,0}+i\ \kk Z_{F,1}+\calo(\kk^2)\right)\,,\\
&\ww=\frac{v}{\sqrt{2}}\ \kk-\frac i2\ \Gamma\ \kk^2+\calo(\kk^3)\,.
\end{split}
\eqlabel{soundq}
\end{equation}
It is straightforward to derive the corresponding equations of motion from \eqref{s1} and \eqref{s2} ---
they are too long to be presented here. We explain the boundary conditions only:
\nxt in the UV, \ie as $x\to 0_+$, 
\begin{equation}
\begin{split}
&Z_{H,0}=x^3+\calo(x^6)\,,\qquad Z_{F,0}=z_{f,0,0}\ x^4 +\calo(x^7)\,,\\
&Z_{H,1}=-\frac{v}{\sqrt{2}}\ x^4+\calo(x^5)\,,\qquad Z_{F,1}=z_{f,1,0}\ x^4 +\calo(x^5)\,;
\end{split}
\eqlabel{soundbc1}
\end{equation}
\nxt in the IR, \ie as $y\equiv 1-x\to 0_+$, 
\begin{equation}
\begin{split}
&Z_{H,0}=z_{h,0,0}^h+\calo(y)\,,\qquad Z_{F,0}=z_{f,0,0}^h+\calo(y)\,,\\
&Z_{H,1}=z_{h,1,0}^h+\calo(y)\,,\qquad Z_{F,1}=z_{f,1,0}^h+\calo(y)\,.
\end{split}
\eqlabel{sound2}
\end{equation}
Given \eqref{soundq}, these asymptotes reflects the incoming boundary conditions
as the horizon, and the Dirichlet boundary conditions at the asymptotic boundary
\cite{Kovtun:2005ev}.

\begin{figure}[t]
\begin{center}
\psfrag{b}[cc][][1.0][0]{$b$}
\psfrag{v}[cc][][1.0][0]{$1-2 c_s^2$}
\psfrag{z}[cc][][1.0][0]{${\zeta}/{\eta}$}
\includegraphics[width=3in]{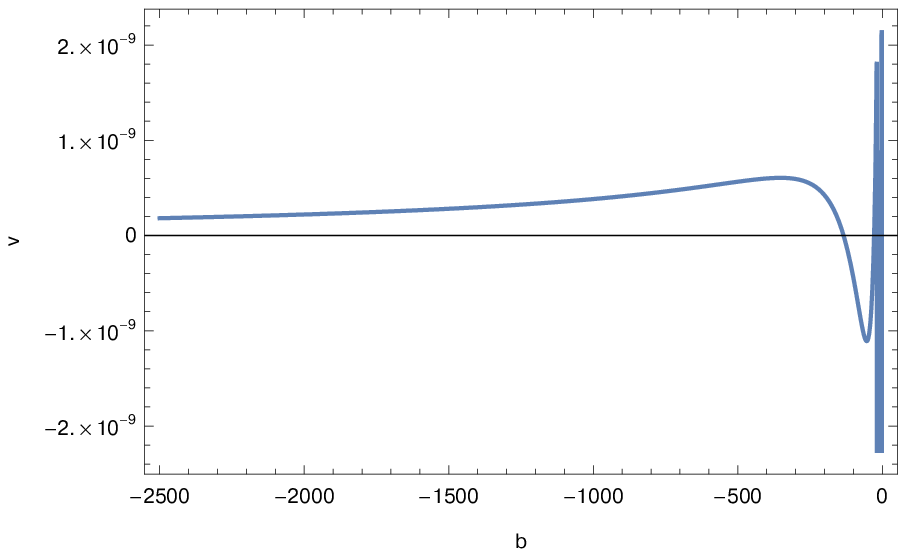}\,
\includegraphics[width=3in]{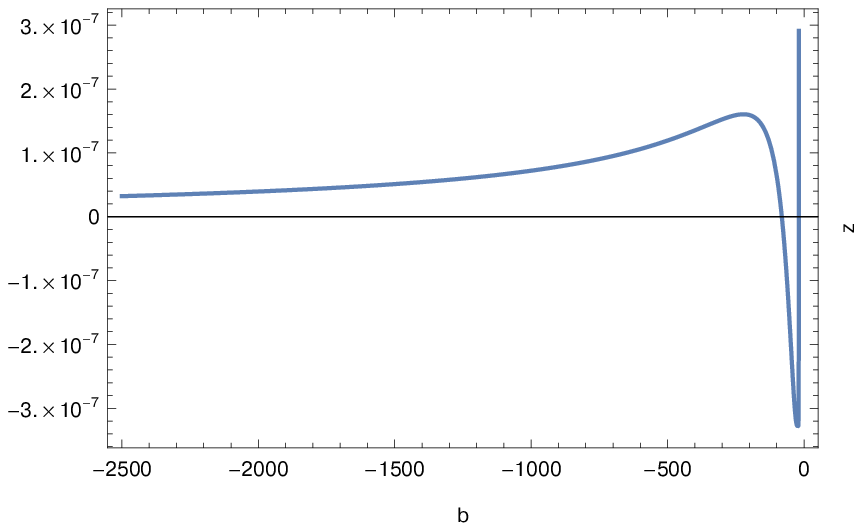}
\end{center}
  \caption{The speed of the sound waves (the left panel) and the bulk viscosity (the right panel)
  of the holographic conformal order as a function of $b$ obtained from the analysis of the sound channel
  QNMs.
} \label{fig3a}
\end{figure}

Numerically solving the equations for $\{Z_{H,0},Z_{F,0},Z_{H,1},Z_{F,1}\}$, subject to the boundary conditions
\eqref{soundbc1} and \eqref{sound2}, we extract
\begin{equation}
1-2 c_s^2=1-v^2\qquad {\rm and}\qquad \frac{\zeta}{\eta}=\Gamma-1\,,
\eqlabel{cszeta2}
\end{equation}
where we used the universal result for the ratio of the shear viscosity to the entropy density 
\cite{Buchel:2003tz}. These results are presented in fig.~\ref{fig3a}. As expected, the 
transport in the ordered phase is conformal. 

\subsubsection{Dispersion of the sound waves in holographic conformal order}\label{dis}

We now present results for the dispersion relation of the sound waves at finite $\kk$ in
holographic conformal order. We set
\begin{equation}
\begin{split}
&\ww(\kk)=\ww_r(\kk)+i\ \ww_i(\kk)\,,\qquad {\rm with}\qquad \lim_{\kk\to 0} \ww(\kk)=0\,,\\
&Z_H=(1-x)^{-i\ww}\ \biggl(Z_{H,r}+i\ Z_{H,i}\biggr)\,,\qquad Z_F=(1-x)^{-i\ww}\ \biggl(Z_{F,r}+i\
Z_{F,i}\biggr)\,,
\end{split}
\eqlabel{defsk}
\end{equation}
and obtain from \eqref{s1} and \eqref{s2} equations\footnote{The equations are too long to
be presented here.} for $\{Z_{H,r},Z_{H,i},Z_{F,r},Z_{F,i}\}$.
These equations have to be solved subject to the boundary conditions: 
\nxt in the UV, \ie as $x\to 0_+$, 
\begin{equation}
\begin{split}
&Z_{H,r}=x^3+\calo(x^4)\,,\qquad Z_{H,i}=-\ww_r\ x^4 +\calo(x^5)\,,\\
&Z_{F,r}=z_{f,r,0}\ x^4+\calo(x^5)\,,\qquad Z_{F,i}=z_{f,i,0}\ x^4 +\calo(x^5)\,;
\end{split}
\eqlabel{soundk1}
\end{equation}
\nxt in the IR, \ie as $y\equiv 1-x\to 0_+$, 
\begin{equation}
\begin{split}
&Z_{H,r}=z_{h,r,0}^h+\calo(y)\,,\qquad Z_{H,i}=z_{h,i,0}^h+\calo(y)\,,\\
&Z_{F,r}=z_{f,r,0}^h+\calo(y)\,,\qquad Z_{F,i}=z_{f,i,0}^h+\calo(y)\,.
\end{split}
\eqlabel{soundk2}
\end{equation}
Given \eqref{defsk}, these asymptotes reflects the incoming boundary conditions
as the horizon, and the Dirichlet boundary conditions at the asymptotic boundary
\cite{Kovtun:2005ev}.

\begin{figure}[t]
\begin{center}
\psfrag{k}[cc][][1.0][0]{$\kk$}
\psfrag{r}[cc][][1.0][0]{$\Re[ \ww]$}
\includegraphics[width=3in]{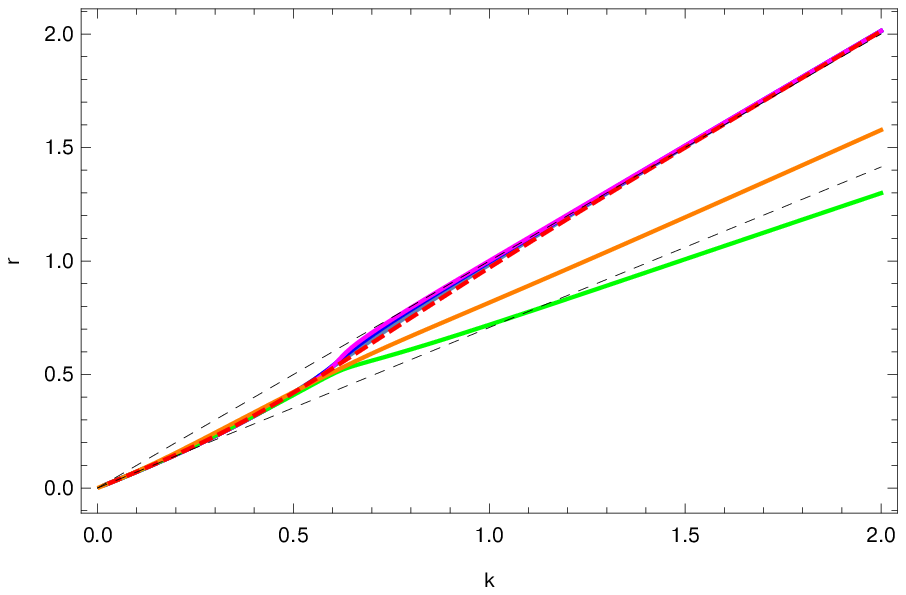}\,
\includegraphics[width=3in]{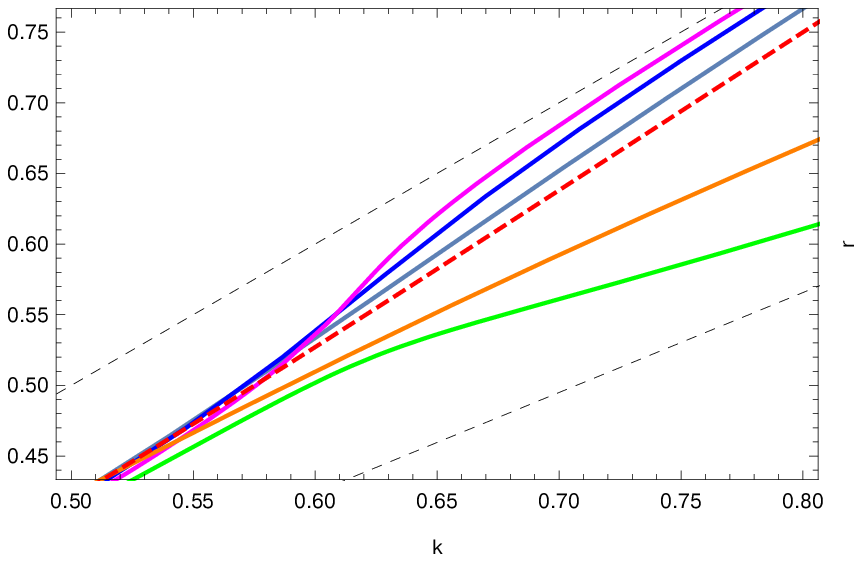}
\end{center}
  \caption{$\Re[\ww]$ of the sound waves dispersion relation in the disordered
  (the dashed red curve) and ordered phases (solid curves) for select values of $b$, see the text below.
  The dashed black lines indicate the hydrodynamic $\kk\to 0$ and the $\kk\to \infty$ limits.
} \label{fig4}
\end{figure}

\begin{figure}[t]
\begin{center}
\psfrag{k}[cc][][1.0][0]{$\kk$}
\psfrag{i}[cc][][1.0][0]{$\Im[ \ww]$}
\includegraphics[width=3in]{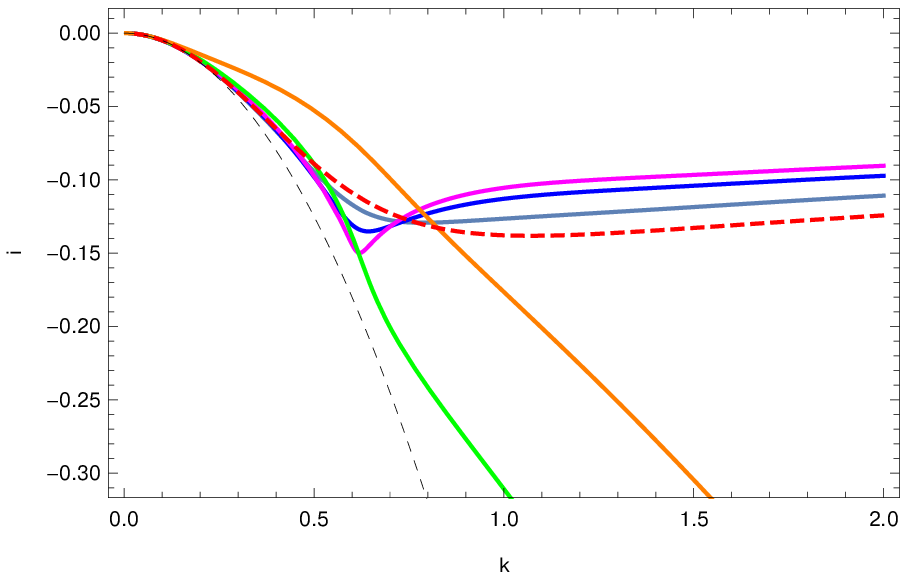}\,
\includegraphics[width=3in]{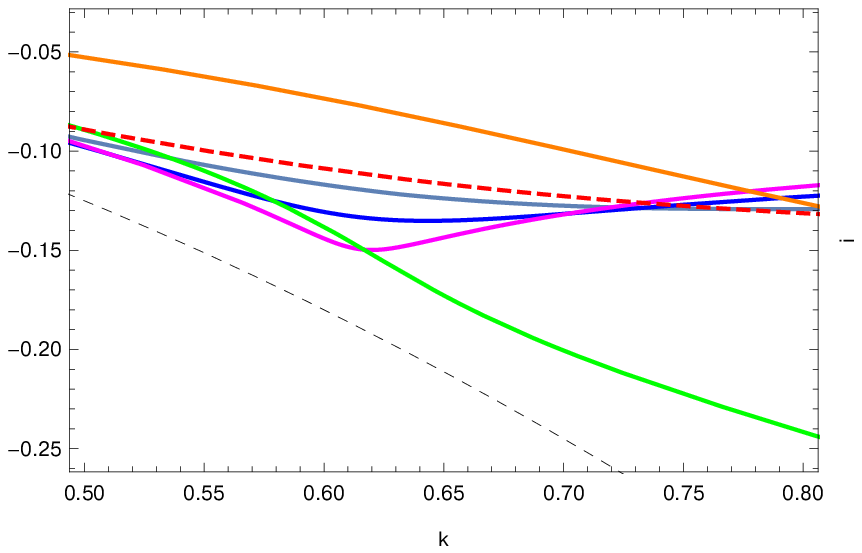}
\end{center}
  \caption{$\Im[\ww]$ of the sound waves dispersion relation in the disordered
  (the dashed red curve) and ordered phases (solid curves) for select values of $b$, see the text below.
  The dashed black line indicates the hydrodynamic $\kk\to 0$ limit.
} \label{fig5}
\end{figure}

In figs.~\ref{fig4}-\ref{fig5} we present results for $\ww_r\equiv \Re[\ww]$ and
$\ww_i\equiv \Im[\ww]$ of the
sound mode in holographic conformal order at $b=\{-10,-5,-4,-3,-2\}$
(the solid $\{$ grey, blue, magenta, green, orange $\}$ curves correspondingly). 
The dashed red curve represents the dispersion relation for the disordered phase,
also in the limit $b\to -\infty$, see section \ref{brancheso}. The dashed black lines
represent the hydrodynamic approximation
\[
\lim_{\kk\to 0}\ \frac{\ww_r(\kk)}{\kk}=\frac{1}{\sqrt 2}\,,\qquad \lim_{\kk\to 0}\ \frac{\ww_i(\kk)}{\kk^2}=-\frac 12\,,
\]
and the large $\kk$ limit, 
\[
\lim_{\kk\to \infty}\ \frac{\ww_r(\kk)}{\kk}=1\,.
\]
There is a noticeable deviation in the sound QNM mode dispersion between the ordered and the disordered
phases for $b\gtrsim -5$; this is a reflection of the fact that in this regime some non-hydrodynamic
modes becoming light, see fig.~\ref{fig7}.

\subsection{Non-hydrodynamic QNMs and the instability}\label{instability}

In this section we discuss the spectrum of low-lying non-hydrodynamic QNMs.
We consider only the non-hydrodynamic modes in the sound channel. 
We identify the QNM with $\Im[\ww_u]>0$, see \eqref{qcrit}, rendering the   
holographic conformal order discussed in this paper unstable. This mode
is present in the spectrum for all values of $b$, whenever the conformal order is present.
We explain why this mode is present in the ordered phases in the limit\footnote{Recall
that in this limit there is no distinction between the ordered and the
disordered phases thermodynamics.} $b\to -\infty$,
and is absent in the
disordered phase.

\begin{figure}[t]
\begin{center}
\psfrag{i}[cc][][1.0][0]{$\Im[ \ww]$}
\psfrag{r}[cc][][1.0][0]{$\Re[ \ww]$}
\includegraphics[width=5in]{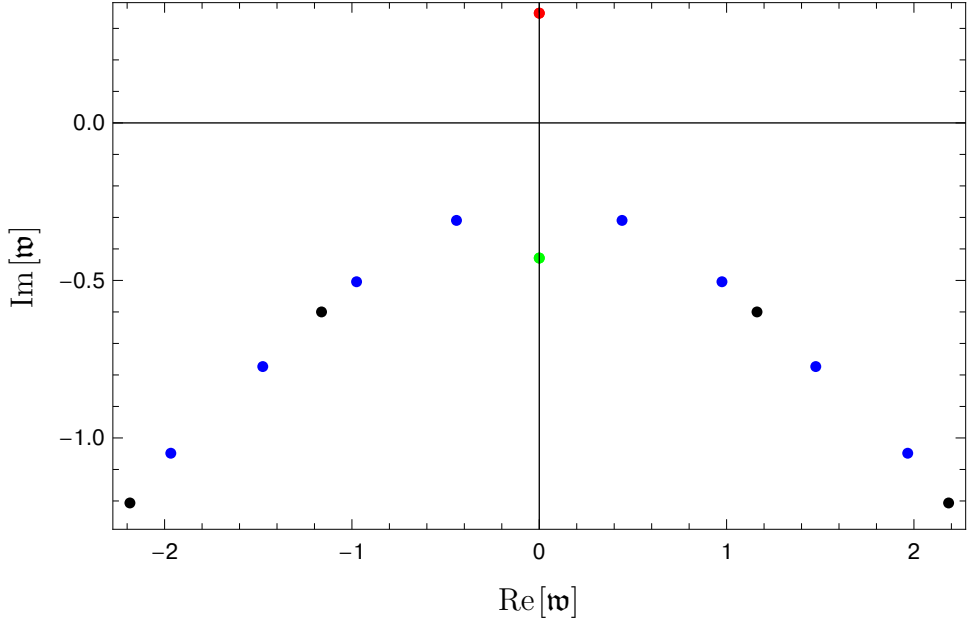}
\end{center}
  \caption{Spectrum of low-lying QNMs in the ordered phase at $b=-4$.
  There are two branches ${\cal{BR}}_{metric}$ (blue dots) and ${\cal{BR}}_{hair}$
  (black dots). The QNMs on the former branch reproduce the QNMs of the $AdS_4$-Schwarzschild
  black brane in the limit $b\to -\infty$, while those on the ${\cal{BR}}_{hair}$ branch
  remain distinct in the limit. QNMs on the ${\cal{BR}}_{hair}^{\Im}$ sub-branch of the
  ${\cal{BR}}_{hair}$ branch are non-propagating: $\Re[\ww]=0$. There is a single dissipative mode
  (a green dot) and the unstable mode (the red dot).
} \label{fig6}
\end{figure}

To begin, we set $\kk=0$. There are two distinct branches of the sound channel QNMs: we call them
$\mathcal{BR}_{metric}$ and $\mathcal{BR}_{hair}$.
\nxt ${\cal{BR}}_{metric}$ branch of the sound channel QNMs. Notice that, see \eqref{ch}, 
\begin{equation}
\calc_{H}\bigg|_{\kk=0}=0\,,
\eqlabel{ch0}
\end{equation}
which decouples $Z_H$ and $Z_F$ fluctuations. On this branch, the spectrum is completely determined
by the equation for  $Z_H$ fluctuations, even though $Z_F$ fluctuations are sourced by the former ones:
$\calb_F\ne 0$ and $\cald_F\ne 0$ in \eqref{s2}. Using the decomposition as in \eqref{defsk},
but for $\lim_{\kk\to 0}\ww(\kk)\ne 0$, we find
\begin{equation}
\begin{split}
&0=Z_{H,r}''+\frac{2 (x^2+x+1) (2 \ww_i x+x-1)-a_3^2 \tV}{2(x^3-1) x}\ Z_{H,r}'
+\frac{2 \ww_r}{x-1}\ Z_{H,i}'-\frac{1}{2x a_1^2 (x^3-1)^2}\\
&\times\biggl(
2 a_3^2 x (\ww_i^2-\ww_r^2) \alpha^2+a_1^2 (x^2+x+1) (a_3^2 \tV \ww_i
-2 (x^2+x+1) ((\ww_i^2-\ww_r^2) x-\ww_i))\biggr)\\
&\times Z_{H,r}
-\frac{\ww_r}{2x a_1^2 (x^3-1)^2} \biggl(
a_3^2 a_1^2 (x^2+x+1) \tV+4 a_3^2 \ww_i x \alpha^2-2 a_1^2 (x^2+x+1)^2 (2 \ww_i x\\
&-1)\biggr)\ Z_{H,i}\,,
\end{split}
\eqlabel{zh1eq}
\end{equation}
\begin{equation}
\begin{split}
&0=Z_{H,i}''-\frac{a_3^2 \tV-2 (x^2+x+1) (2 \ww_i x+x-1)}{2(x^3-1) x}\ Z_{H,i}'
-\frac{2 \ww_r}{x-1}\ Z_{H,r}'-\frac{1}{2x a_1^2 (x^3-1)^2}\\&
\times\biggl(
a_1^2 a_3^2 \ww_i (x^2+x+1) \tV+2 a_3^2 x (\ww_i^2-\ww_r^2) \alpha^2
-2 a_1^2 (x^2+x+1)^2 ((\ww_i^2-\ww_r^2) x-\ww_i)\biggr)
\\&\times Z_{H,i}+\frac{\ww_r}{2x a_1^2 (x^3-1)^2} \biggl(a_1^2 a_3^2 (x^2+x+1) \tV+4 a_3^2 \ww_i x \alpha^2
-2 a_1^2 (x^2+x+1)^2 (2 \ww_i x-1)\\
&\biggr)\ Z_{H,r}\,.
\end{split}
\eqlabel{zh2eq}
\end{equation}
The boundary conditions are as in the first lines of eq.~\eqref{soundk1} and \eqref{soundk2}.
Note that the limit $b\to -\infty$ is trivial here (see \eqref{tildedef} and \eqref{perlimit}):
\begin{equation}
\tV\to -6\,,\qquad a_1\to 1\,,\qquad a_3\to 1\,,\qquad \alpha\to 3\,,
\eqlabel{blimitzh}
\end{equation}
precisely reproducing the equations for the $\kk=0$ non-hydrodynamic
QNMs of the $AdS_4$ Schwarzschild black brane ---  there is no distinction between
the ordered and the disordered QNMs in this limit. We present the spectrum of the quasinormal
modes on this branch for $b=-4$ as blue dots in fig.~\ref{fig6}. 
\nxt ${\cal{BR}}_{hair}$ branch of the sound channel QNMs. Setting identically $Z_H\equiv 0$,
in addition to $\kk=0$, we satisfy the QNM equation \eqref{s1}.  Using the
decomposition as in \eqref{defsk}  we find
\begin{equation}
\begin{split}
&0=Z_{F,r}''-\frac{a_3^2 \tV-2 (x^2+x+1) (2 \ww_i x+x-1)}{2(x^3-1) x}\ Z_{F,r}'
+\frac{2 \ww_r}{x-1}\ Z_{F,i}'+\frac{1}{4a_1^2 x^2 (x^3-1)^2}\\
&\times\biggl(
4 a_3^2 a_1^2 (x^3-1) \hred{\del^2\tV}-4 \chi'  a_3^2 a_1^2 x (x^3-1) \del\tV 
+a_3^2 a_1^2 x (x^2+x+1)( (\chi')^2 x (x-1)\\&-2 \ww_i)
\tV-4 a_3^2 x^2 (\ww_i^2-\ww_r^2)
\alpha^2+4 a_1^2 (x^2+x+1) (4 a_3^2 (x-1)+x (x^2+x+1)\\&\times
((\ww_i^2-\ww_r^2) x-\ww_i))\biggr)\ Z_{F,r}
-\frac{\ww_r}{2x a_1^2 (x^3-1)^2} (a_1^2 a_3^2 (x^2+x+1) \tV+4 a_3^2 \ww_i x \alpha^2\\
&-2 a_1^2
(x^2+x+1)^2 (2 \ww_i x-1)\biggr)\ Z_{F,i}\,,
\end{split}
\eqlabel{zf1eq}
\end{equation}
\begin{equation}
\begin{split}
&0=Z_{F,i}''-\frac{a_3^2 \tV-2 (x^2+x+1) (2 \ww_i x+x-1)}{2x (x^3-1)}\  Z_{F,i}'
-\frac{2 \ww_r}{x-1}\ Z_{F,r}'+\frac{1}{4a_1^2 x^2 (x^3-1)^2}
\\& \times \biggl(
4 a_3^2 a_1^2 (x^3-1) \hred{\del^2\tV}-4 \chi' a_3^2 a_1^2 x (x^3-1) \del\tV
-4 a_3^2 x^2 (\ww_i^2-\ww_r^2) \alpha^2+a_1^2 (x^2+x+1)\\
&\times (a_3^2 \tV x (\chi')^2 x (x-1)
-2 \ww_i)+16 a_3^2 (x-1)+4 x (x^2+x+1) ((\ww_i^2-\ww_r^2) x-\ww_i)\biggr) Z_{F,i}\\
&+\frac{\ww_r}{2x a_1^2 (x^3-1)^2}
\biggr(a_1^2 a_3^2 (x^2+x+1) \tV+4 a_3^2 \ww_i x \alpha^2-2 a_1^2 (x^2+x+1)^2 (2 \ww_i x-1)\biggr) Z_{F,r}\,.
\end{split}
\eqlabel{zf2eq}
\end{equation}
The boundary conditions are
\begin{equation}
Z_{F,r}=x^4+\calo(x^5)\,,\qquad Z_{F,i}=-\ww_r\ x^5+\calo(x^6)\,,
\eqlabel{bcfuv}
\end{equation}
as $x\to 0_+$, and 
\begin{equation}
Z_{F,r}=z_{f,r,0}^h+\calo(y)\,,\qquad Z_{F,i}=z_{f,i,0}^h+\calo(y)\,,
\eqlabel{bcfir}
\end{equation}
as $y\equiv 1-x\to 0_+$. 
Now, there is a clear distinction between the ordered and the disordered phases, even in the limit
$b\to -\infty$. Indeed, the QNM equations \eqref{zf1eq} and \eqref{zf2eq} contain $\del^2\tV$
(highlighted in red), and
\begin{equation}
\lim_{b\to -\infty} \del^2\tV=\lim_{b\to -\infty} \left(4+12 b\chi^2\ \right)=4-12 \left(f_{[1]}(x)\right)^2\,,
\eqlabel{deltvlimit}
\end{equation}
where we used \eqref{perlimit}. In the ordered phase $f_{[1]}(x)$ is a nontrivial function,
see fig.~\ref{fig1}, while in the disordered phase it vanishes identically.  We present the spectrum
of the quasinormal modes on this branch for $b=-4$ as black dots in fig.~\ref{fig6}. 
\nxt There is a non-propagating sub-branch of  ${\cal{BR}}_{hair}$ branch, we call it
 ${\cal{BR}}_{hair}^{\Im}$.
QNM equation of motion on this sub-branch is a consistent truncation of
eqs.~\eqref{zf1eq} and \eqref{zf2eq} 
with
\begin{equation}
\ww_r=0\,,\qquad Z_{F,i}\equiv 0\,,
\eqlabel{qnmtruncation}
\end{equation}
resulting in 
\begin{equation}
\begin{split}
&0=Z_{F,r}''-\frac{a_3^2 \tV-2 (x^2+x+1) (2 \ww_i x+x-1)}{2(x^3-1) x}\  Z_{F,r}'+\frac{2 \ww_r}{x-1}\
Z_{F,i}'+\frac{1}{4a_1^2 x^2 (x^3-1)^2}\\&\times
\biggl(4 a_3^2 a_1^2 (x^3-1) \del^2\tV-4 \chi' a_3^2 a_1^2 x (x^3-1) \del\tV
+a_3^2 a_1^2 x (x^2+x+1) ((\chi')^2 x (x-1)\\
&-2 \ww_i) \tV-4 a_3^2 x^2 (\ww_i^2-\ww_r^2)
\alpha^2+4 a_1^2 (x^2+x+1) (4 a_3^2
(x-1)+x (x^2+x+1)
\\
&\times((\ww_i^2-\ww_r^2) x-\ww_i))\biggr)\ Z_{F,r}-\frac{\ww_r}{2x a_1^2 (x^3-1)^2} \biggl(
a_1^2 a_3^2 (x^2+x+1) \tV+4 a_3^2 \ww_i x \alpha^2\\&-2 a_1^2 (x^2+x+1)^2 (2 \ww_i x-1)\biggr)\ Z_{F,i}\,.
\end{split}
\eqlabel{zfeq}
\end{equation}
Solving \eqref{zfeq} with the boundary conditions for $Z_{F,r}$ as in \eqref{bcfuv} and \eqref{bcfir},
we find two quasinormal modes: one with $\ww_i=\Im[\ww]<0$ and the other one with $\ww_i=\Im[\ww]>0$ ---
these are (correspondingly) the green and the red dots presented for $b=-4$ in fig.~\ref{fig6}.
The red dot QNM is what we called $\ww_u$ in section \ref{intro}; it is signaling perturbative
instability of the hairy black brane horizon, dual to a holographic conformal order.

In the rest of this section we focus on the  ${\cal{BR}}_{hair}^{\Im}$ branch.
First, we identify the ${\cal{BR}}_{hair}^{\Im}$ branch at $\kk\ne 0$. 
Using the decomposition \eqref{defsk}, we find that the truncation
\begin{equation}
\ww_r=0\,,\qquad Z_{F,i}\equiv 0\,,\qquad Z_{H_i}\equiv 0\,,
\eqlabel{trunc2}
\end{equation}
is a consistent one, even for $\kk\ne 0$:
\begin{equation}
\begin{split}
&0=Z_{F,r}''+\cala_F^{\Im}\ Z_{F,r}'+\calb_F^{\Im}\ Z_{H,r}' +\calc_F^{\Im}\ Z_{F,r}+\cald_F^{\Im}\ Z_{H,r}\,,\\
&0=Z_{H,r}''+\cala_H^{\Im}\ Z_{H,r}'+\calb_H^{\Im}\ Z_{H,r} +\calc_H^{\Im}\ Z_{F,r}\,,
\end{split}
\eqlabel{uns1}
\end{equation}
with
\begin{equation}
\begin{split}
&\cala_F^{\Im}=-\frac{a_3^2 \tV-2 (x^2+x+1) (2 \ww_i x+x-1)}{2(x^3-1) x}\,,
\end{split}
\eqlabel{afun}
\end{equation}
\begin{equation}
\begin{split}
&\calb_F^{\Im}=-\frac{2 a_3^2 \ww_i^2 (\chi' x \tV-2 \del\tV)}{(a_1^2 \kk^2 x^2 (\chi')^2 (x^3-1)+2 a_1^2 \kk^2 (a_3^2 \tV+2 x^3-2)-16 \ww_i^2) (x^3-1) x}\,,
\end{split}
\eqlabel{bfun}
\end{equation}
\begin{equation}
\begin{split}
&\calc_F^{\Im}=-\frac12 \biggl(
a_3^2 a_1^4 \del\tV \kk^2 x^3 (x^3-1)^2 (\chi')^3+a_1^2 x^2 (x^3-1)
(-2 a_3^2 a_1^2 \kk^2 (x^3-1) \del^2\tV+a_3^2 (a_1^2 \kk^2\\
&\times(x^2+x+1)
(\ww_i x-8 x+8)+8 \ww_i^2) \tV-2 \kk^2 a_3^2 x^2 (a_1^2 \kk^2 (x^3-1)-\ww_i^2)
\alpha^2-2 a_1^2 \kk^2\\&\times
(x^2+x+1) (x \ww_i (x^2+x+1) (\ww_i x-1)+4 a_3^2 (x-1)))
(\chi')^2+2 a_3^2 a_1^2 \del\tV x (x^3-1)\\
&\times
(a_1^2 \kk^2 (a_3^2 \tV+10 x^3-10)-16 \ww_i^2) \chi'-4 a_3^2 a_1^2 (x^3-1) (a_1^2 \kk^2 (a_3^2 \tV+2 x^3-2)-8 \ww_i^2)\\
&\times \del^2\tV-4 a_3^2 x^2 (a_1^2 \kk^2 (x^3-1)-\ww_i^2) (a_1^2 \kk^2 (a_3^2 \tV+2 x^3-2)-8 \ww_i^2)
\alpha^2-4 a_1^2 a_3^2\\
&\times (x^2+x+1) (a_1^2 \kk^2
(x^2 \ww_i(x^2+x+1) (\ww_i-1)+4 a_3^2 (x-1))+4 \ww_i^3 x) \tV+2 a_1^4 a_3^4 \kk^2 \ww_i\\&
\times x (x^2+x+1) \tV^2
-8 a_1^2 (x^2+x+1) (x \ww_i (x^2+x+1) (\ww_i x-1)+4 a_3^2 (x-1)) (a_1^2 \kk^2 \\
&\times (x^3-1)-4 \ww_i^2)\biggr)
\biggl(a_1^2 x^2 (x^3-1)^2 (a_1^2 \kk^2 x^2 (\chi')^2 (x^3-1)+2 a_1^2 \kk^2 (a_3^2 \tV+2 x^3
-2)\\
&-16 \ww_i^2)\biggr)^{-1}\,,
\end{split}
\eqlabel{cfun}
\end{equation}
\begin{equation}
\begin{split}
&\cald_F^{\Im}=\frac14 \biggl(
(\chi')^2 x^2 (x^3-1)+2 a_3^2 \tV+4 (x^2+x+1) (2 \ww_i x-3 x+3)\biggr)
\biggl(
2 \del\tV-\chi' x \tV\biggr)\\
&\times\ww_i^2 a_3^2\biggl(
(a_1^2 \kk^2 x^2 (\chi')^2 (x^3-1)
+2 a_1^2 \kk^2 (a_3^2 \tV+2 x^3-2)-16 \ww_i^2) (x^3-1)^2 x^2\biggr)^{-1}\,,
\end{split}
\eqlabel{dfun}
\end{equation}
\begin{equation}
\begin{split}
&\cala_H^{\Im}=-\frac12 \biggl(
-a_1^2 \kk^2 x^4 (x^3-1)^2 (\chi')^4-a_1^2 \kk^2 x^2 (x^3-1) (a_3^2 \tV+2 (x^2+x+1)
(2 \ww_i x-3 x\\
&+3)) (\chi')^2-16 \ww_i^2 (a_3^2 \tV-2 (x^2+x+1) (2 \ww_i x+x-1))
+2 a_1^2 (a_3^4 \tV^2-4 \tV (x^2+x+1)\\
&\times(\ww_i x+x-1) a_3^2-4 (x-1)
(x^2+x+1)^2 (2 \ww_i x-5 x+5)) \kk^2\biggr)\biggl(
(a_1^2 \kk^2 x^2 (\chi')^2
(x^3-1)\\
&+2 a_1^2 \kk^2 (a_3^2 \tV+2 x^3-2)-16 \ww_i^2) (x^3-1) x\biggr)^{-1}\,,
\end{split}
\eqlabel{ahun}
\end{equation}
\begin{equation}
\begin{split}
&\calb_H^{\Im}=\frac{1}{16} \biggl(
a_1^4 \kk^2 x^6 (x^3-1)^3 (\chi')^6+4 a_1^4 \kk^2 x^4 (x^3-1)^2 (a_3^2 \tV+(x^2+x+1) (2 \ww_i x-5 x\\
&+5)) (\chi')^4+4 a_1^2 \kk^2 x^2 (x^3-1) (a_3^4 a_1^2 \tV^2+2 a_3^2 a_1^2 (x^2+x+1)
(\ww_i x-4 x+4) \tV+4 a_3^2 x^2\\&
\times(a_1^2 \kk^2 (x^3-1)-\ww_i^2) \alpha^2+4 a_1^2 (x^2+x+1)^2 (\ww_i^2 x^2-4 \ww_i x^2+3 \ww_i x+3 x^2-6 x+3))\\
&\times (\chi')^2+32 a_3^2 x^2 (a_1^2 \kk^2 (x^3-1)-\ww_i^2) (a_1^2 \kk^2 (a_3^2 \tV+2 x^3-2)-8 \ww_i^2) \alpha^2-16 a_1^2 (x^2+x\\
&+1) (a_3^4 a_1^2 \kk^2 (\ww_i x-x+1) \tV^2-2 a_3^2 (a_1^2 \kk^2 (x^2+x+1) (\ww_i^2 x^2+\ww_i x^2-2 \ww_i x-6 x^2+12 x\\
&-6)+4 \ww_i^3 x) \tV-4 (x^2+x+1) (a_1^2 \kk^2 (x^3-1) (\ww_i^2 x^2-6 \ww_i x^2+5 \ww_i x+9 x^2-18 x+9)
\\&+4 \ww_i^3 x-4 \ww_i^4 x^2))\biggl)
\biggl(
a_1^2 x^2 (x^3-1)^2 (a_1^2 \kk^2 x^2 (\chi')^2
(x^3-1)+2 a_1^2 \kk^2 (a_3^2 \tV+2 x^3-2)\\&-16 \ww_i^2)\biggl)\,,
\end{split}
\eqlabel{bhun}
\end{equation}
\begin{equation}
\begin{split}
&\calc_H^{\Im}=-\frac14 a_1^2 \kk^2 \biggl(
(\chi')^2 x^2 (x^3-1)+2 a_3^2 \tV-12 x^3+12\biggr)
\biggl(-4 \ww_i^2 x^3 (x^3-1) (\chi')^3+a_3^2 a_1^2
\del\tV\\&\times  \kk^2 x^2 (x^3-1) (\chi')^2
-8 x (x^3-1) (a_3^2 a_1^2 \kk^2 \tV-6 \ww_i^2) \chi'+2 a_3^2 \del\tV (a_1^2 \kk^2 (a_3^2 \tV+2 x^3-2)\\
&-8 \ww_i^2)\biggr)
\biggl(
(a_1^2 \kk^2 x^2 (\chi')^2 (x^3-1)+2 a_1^2 \kk^2
(a_3^2 \tV+2 x^3-2)-16 \ww_i^2) (x^3-1) x^2 \ww_i^2\biggr)\,.
\end{split}
\eqlabel{chun}
\end{equation}
Eqs.~\eqref{uns1} have to be solved subject to the following boundary conditions:
\nxt in the UV, \ie as $x\to 0_+$,
\begin{equation}
Z_{F,r}=x^4+\calo(x^5)\,,\qquad Z_{H,r}=z_{h,r,0}\ x^3+\calo(x^4)\,;
\eqlabel{unquv}
\end{equation}
\nxt in the IR, \ie as $y\equiv 1-x\to 0_+$,
\begin{equation}
Z_{F,r}=z_{f,r,0}^h+\calo(y)\,,\qquad Z_{H,r}=z_{h,r,0}^h+\calo(y)\,.
\eqlabel{unqir}
\end{equation}

\begin{figure}[t]
\begin{center}
\psfrag{q}[cc][][1.0][0]{$\kk$}
\psfrag{i}[cc][][1.0][0]{$\Im[\ww_u]$}
\includegraphics[width=5in]{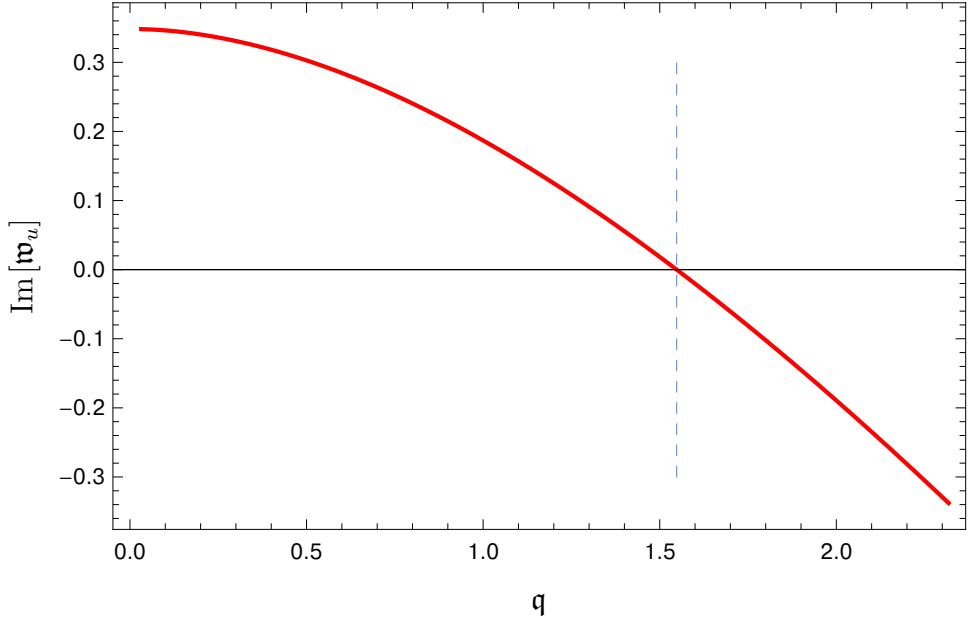}
\end{center}
  \caption{The dispersion relation $\ww_u(\kk)$ of the unstable QNM in the ordered phase
  at $b=-4$. For this mode $\Re[\ww_u]=0$. As typical for the Gregory-Laflamme instability
  \cite{Gregory:1993vy},
  this mode is stabilized for $\kk>\kk_c$ (here represented by a vertical dashed line,
  see eq.~\eqref{qcbm4}.)
} \label{fig9}
\end{figure}

In fig.~\ref{fig9} we present $\ww_u(\kk)$ --- the dispersion relation of 
the unstable QNM in the ordered phase at $b=-4$; this is the $\kk$ dependence of the
red dot in fig.~\ref{fig6}. Note the existence of the critical momenta
\begin{equation}
\kk_c\bigg|_{b=-4}=1.548\,,
\eqlabel{qcbm4}
\end{equation}
represented by vertical dashed line, such that for
\begin{equation}
\kk\ >\ \kk_c\qquad \Longrightarrow\qquad \Im[\ww(\kk)]\ <\ 0\,,
\eqlabel{becsta}
\end{equation}
\ie this QNM becomes stable.

\begin{figure}[t]
\begin{center}
\psfrag{i}[cc][][1.0][0]{$\Im[ \ww]$}
\psfrag{b}[cc][][1.0][0]{$b$}
\includegraphics[width=5in]{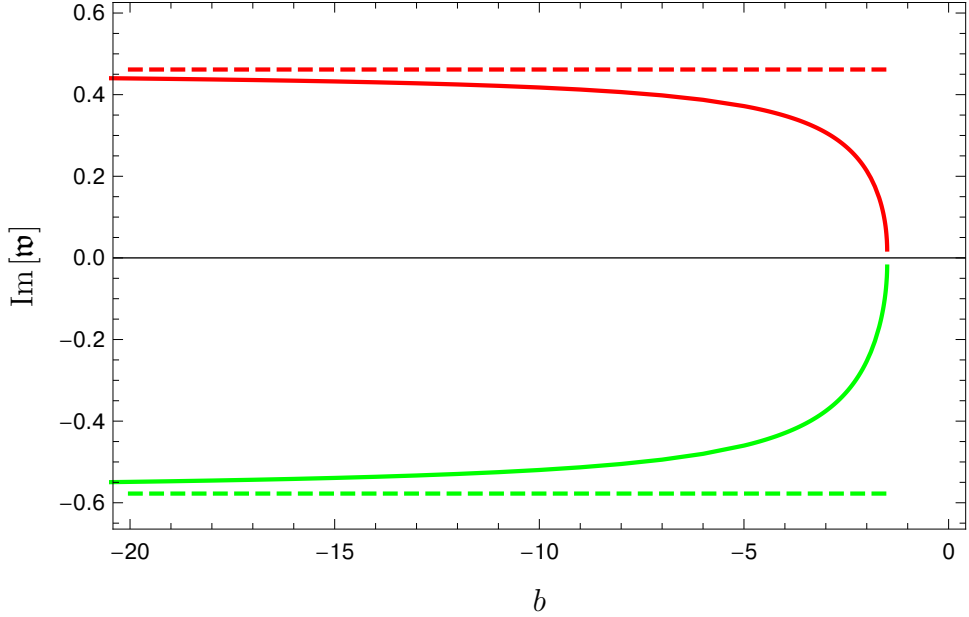}
\end{center}
  \caption{$\ww(\kk=0)$ of the non-propagating ($\Re[\ww]=0$) QNMs on the ${\cal{BR}}_{hair}^{\Im}$ branch
  as a function of $b$ in the ordered phase. The dashed horizontal lines represent the limit
  as $b\to -\infty$, see eq.~\eqref{limitsqnm}. As $b\to b_{crit,0}=-\frac 32$, these QNMs become light,
  strongly affecting the dispersion of the sound waves, see figs.~\ref{fig4}-\ref{fig5}.
} \label{fig7}
\end{figure}

In fig.~\ref{fig7} we present results for $\ww(\kk=0)$ as a function of
$b$ for the QNMs on the ${\cal{BR}}_{hair}^{\Im}$ branch. There is always
an instability in the holographic conformal order, irrespective of the value of
$b$ (the solid red curve). The solid green curve represents the stable QNM.
Both QNMs become light in the limit $b\to b_{crit,0}=-\frac 32$,
strongly affecting the sound waves dispersion, see section \ref{dis}. The dashed horizontal
lines represent the $b\to -\infty$ limit of $\ww$,
\begin{equation}
\lim_{b\to -\infty}
\begin{cases}
 \Im[\ww(0)]=0.461\,,\qquad {\rm unstable\ QNM\ (red)}\,;\\
 \Im[\ww(0)]=-0.577\,,\qquad {\rm stable\ QNM\ (green)}\,.
\end{cases}
\eqlabel{limitsqnm}
\end{equation}

The spectral results \eqref{limitsqnm} are obtained solving eq.~\eqref{zfeq} in the limit
$b\to -\infty$. Taking
\begin{equation}
Z_{F,r}=Z_{F,0}+\calo\left(\frac 1b\right)\,,\qquad \ww_i=\ww_{i,0}+\calo\left(\frac 1b\right)\,,
\eqlabel{expandzfr}
\end{equation}
and using \eqref{perlimit} we find the limiting QNM equation
\begin{equation}
\begin{split}
&0=Z_{F,0}''+\frac{4 \ww_{i,0} x (x^2+x+1)-12+18 x^3}{2x (x^3-1)}\ Z_{F,0}'
-\frac{1}{2x^2 (x^3-1)^2} \biggl( 24 (x^3-1) \hred{f_{[1]}^2}-2 x\\
&\times(x-1) (x (x+2) (x^2+x+4) \ww_{i,0}^2+(x^2+x+1) ((8 x^2+7 x+6) \ww_{i,0}+16 x^2))\biggr)\  Z_{F,0}\,,
\end{split}
\eqlabel{zfeqpert}
\end{equation}
where we highlighted again the crucial difference between the ordered and the disordered
phases:
\begin{equation}
f_{[1]}(x)=\begin{cases}
&\equiv 0\,, \qquad {\rm in\ the\ disordered\ phase}\,;\\
&{\rm nontrivial,\ given\ by\ eq.~}\eqref{peq1}\,, \qquad {\rm in\ the\ ordered\ phase}\,.
\end{cases}
\eqlabel{f1ordis}
\end{equation}

\begin{figure}[t]
\begin{center}
\psfrag{x}[cc][][1.0][0]{$x$}
\psfrag{u}[cc][][1.0][0]{$U_0(x)$}
\psfrag{v}[cc][][1.0][0]{$U_{dis}(x)$}
\includegraphics[width=3in]{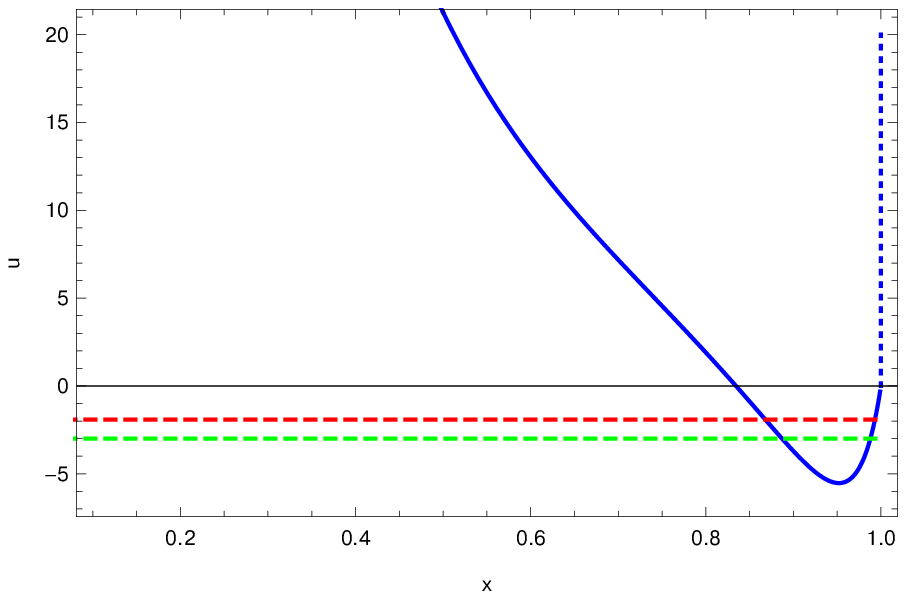}
\includegraphics[width=3in]{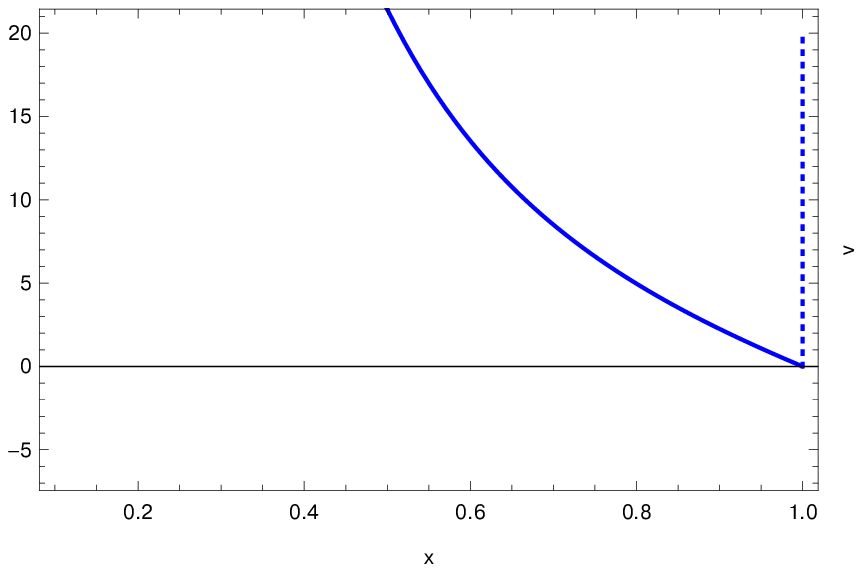}
\end{center}
  \caption{Existence of the unstable QNM as $b\to -\infty$ is equivalent to the existence of the
  bound state in an effective one-dimensional potential \eqref{defux} with $E<0$.
  We present this potential, $U_0(x)$,  for the ordered phase of the overtone 0
  (the left panel). The horizontal dashed lines indicate the effective energies of the
  QNMs \eqref{limitsqnm}. In the right panel we present the effective potential in the
  disordered phase, $U_{dis}$. Since $U_{dis}(x)\ge 0$ for $x\in (0,1]$,
  there can not be any unstable QNM. 
} \label{fig8}
\end{figure}

The asymptotic representation \eqref{zfeqpert} explains why the ordered phase is
always unstable, while the disordered phase is not. Indeed, 
introducing a new radial coordinate $u$ with
\begin{equation}
\frac{du}{dx}=\frac{1}{1-x^3}\,,
\eqlabel{defdudx}
\end{equation}
and rescaling
\begin{equation}
Z_{F,0}=\frac{1}{x^3 (1-x)^{\ww_{i,0}}}\ \Psi_{F}\,,
\eqlabel{rescalez}
\end{equation}
we obtain a Schr\"odinger-like equation for $\Psi_F$:
\begin{equation}
\biggl[-\frac{d^2}{du^2}+U\biggr]\ \Psi_F\ =\ E\ \Psi_F\,,
\eqlabel{schr}
\end{equation}
where the effective 1d potential $U$ is 
\begin{equation}
U(x)=\frac{1-x^3}{x^2}\ \biggl(x^3+6-12 (f_{[1]})^2\biggr)\,,\qquad x\in (0,1)\,,
\eqlabel{defux}
\end{equation}
and
\begin{equation}
E=-9\ \ww_{i,0}^2\,.
\eqlabel{defesch}
\end{equation}
For an unstable QNM, \ie for $\ww_{i,0}>0$, the boundary conditions for $Z_{F,0}$,
\ie $Z_{F,0}\propto x^4$ as $x\to 0_+$ and $Z_{F_,0}\propto (1-x)^0$ as $y=1-x\to 0_+$,
imply that given the definition \eqref{schr},
\begin{equation}
\Psi_F\to 0\,,\qquad {\rm both\ as}\qquad x\to 0_+\qquad  {\rm  and}\qquad  x\to 1_-\,.
\eqlabel{bcpsi}
\end{equation}
The potential $U(x)$ of \eqref{defux} is divergent as $x\to 0_+$, automatically enforcing
the first of the boundary conditions in \eqref{bcpsi};
it is vanishing as $x\to 1_-$, so to enforce  the second boundary condition in
\eqref{bcpsi} we put an infinite domain wall at $x=1$.
Phrased in the language of the effective 1d Schr\"odinger problem with \eqref{schr} and
\eqref{defesch}, the presence of the unstable QNM in the spectrum is equivalent to
the existence of the bound state in the effective potential \eqref{defux} with 
with $E<0$. In fig.~\ref{fig8} we plot the potential \eqref{defux} for the ordered phase, \ie
with $f_{[1]}(x)\ne 0$
(the left panel) and for the disordered phase, \ie with  $f_{[1]}(x)\equiv 0$ (the
right panel). In the former case the potential dips below zero, allowing for the
bound states (represented by the dashed horizontal lines for the QNM frequencies
\eqref{limitsqnm} ), while in the latter case the potential is always non-negative
for $x\in (0,1]$, thus excluding the instability.

\begin{figure}[t]
\begin{center}
\psfrag{b}[cc][][1.0][0]{$\rm branch\ index$}
\psfrag{i}[cc][][1.0][0]{$\Im[\ww]$}
\includegraphics[width=5in]{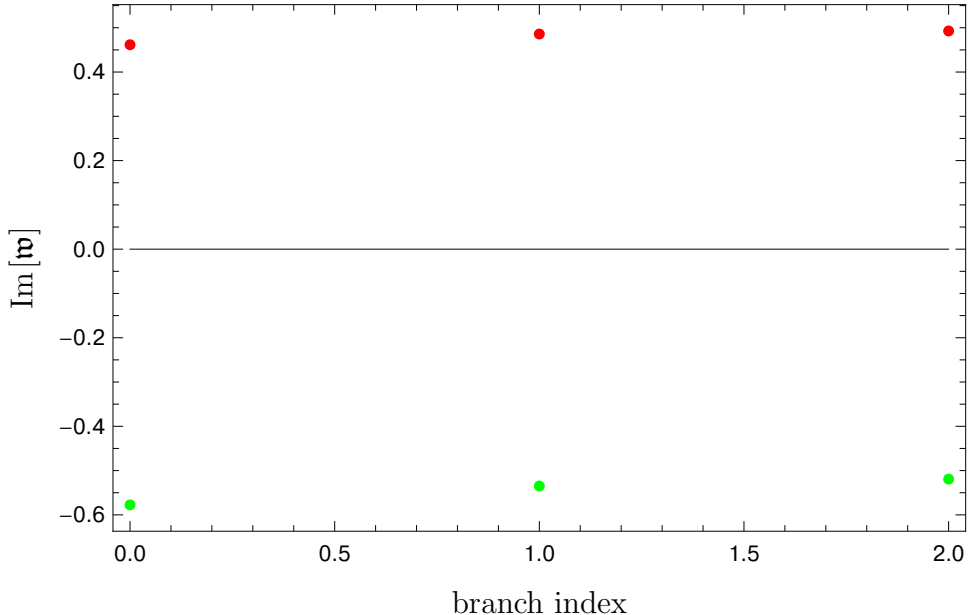}
\end{center}
  \caption{The spectrum of the QNMs on the  ${\cal{BR}}_{hair}^{\Im}$ branch at $\kk=0$ and
  in the limit $b\to -\infty$ for index $i=0,1,2$ overtones of the holographic conformal
  order, see fig.~\ref{fig1}. The higher overtones are more unstable as the value of $\Im[\ww_u]$
  grows with index (red dots).
} \label{fig9a}
\end{figure}

We demonstrated above that there is a perturbative instability on
an index 0 ordered phase branch/overtone for any value of $b\in (-\infty,-\frac 32)$.
In fact, there is an instability on the excited branches/overtones of the
conformal order. In fig.~\ref{fig9a} we present the spectrum of the QNMs on the
${\cal{BR}}_{hair}^{\Im}$ branch for the index $i=0,1,2$ background phase overtones in the
limit $b\to -\infty$. These results are obtained solving \eqref{zfeqpert}
with the bulk scalar profile function $f_{[1]}$ corresponding to the overtone
index $i=0,1,2$, see the left panel of fig.~\ref{fig1}. Note that the higher 
branches of the conformal order are more unstable as the value of $\Im[\ww_u]$ increases
with the overtone index (the red dots).

\section{Numerical tests}

\begin{figure}[t]
\begin{center}
\psfrag{k}[cc][][1.0][0]{$\kk$}
\psfrag{i}[cc][][1.0][0]{$\Im[ \ww]$}
\psfrag{r}[cc][][1.0][0]{$\Re[ \ww]$}
\includegraphics[width=3in]{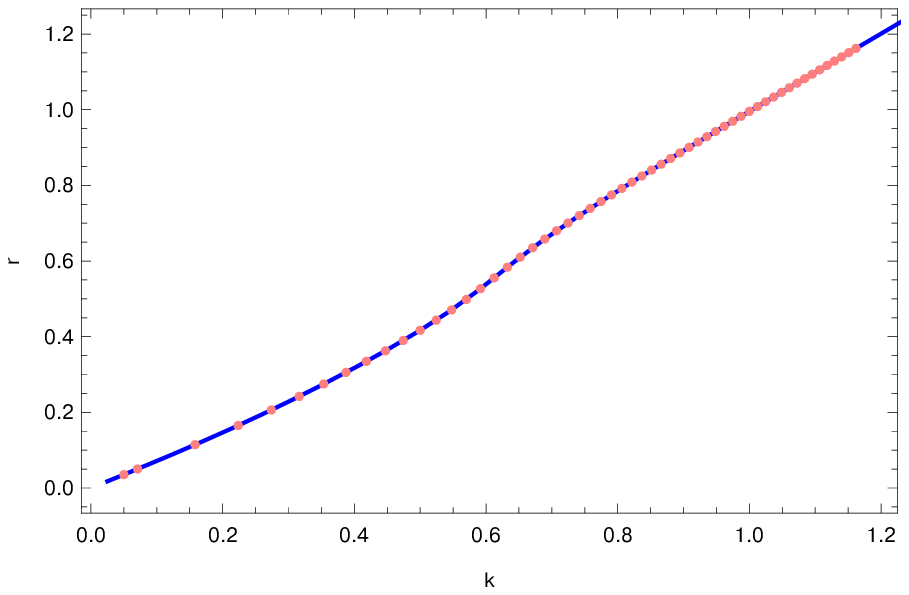}\,
\includegraphics[width=3in]{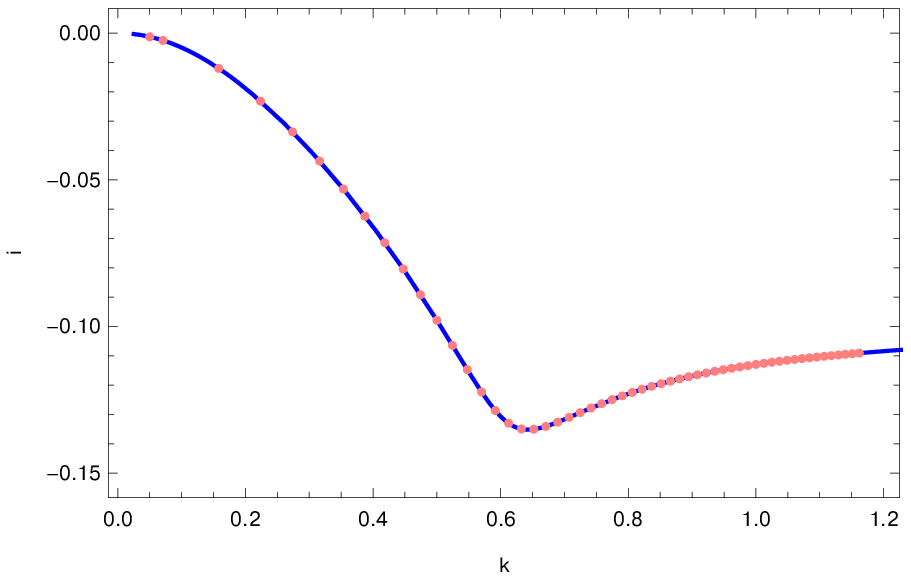}
\end{center}
  \caption{Comparison of the sound waves dispersion in the ordered phase at $b=-5$ extracted using the
  the $x$ radial coordinate (solid curves) and the $\hx$ radial coordinate, see \eqref{relatexxxhx},
  (the dots). 
} \label{figa1}
\end{figure}

\begin{figure}[t]
\begin{center}
\psfrag{k}[cc][][1.0][0]{$\kk$}
\psfrag{i}[cc][][1.0][0]{$\Im[ \ww]$}
\psfrag{r}[cc][][1.0][0]{$\Re[ \ww]$}
\includegraphics[width=3in]{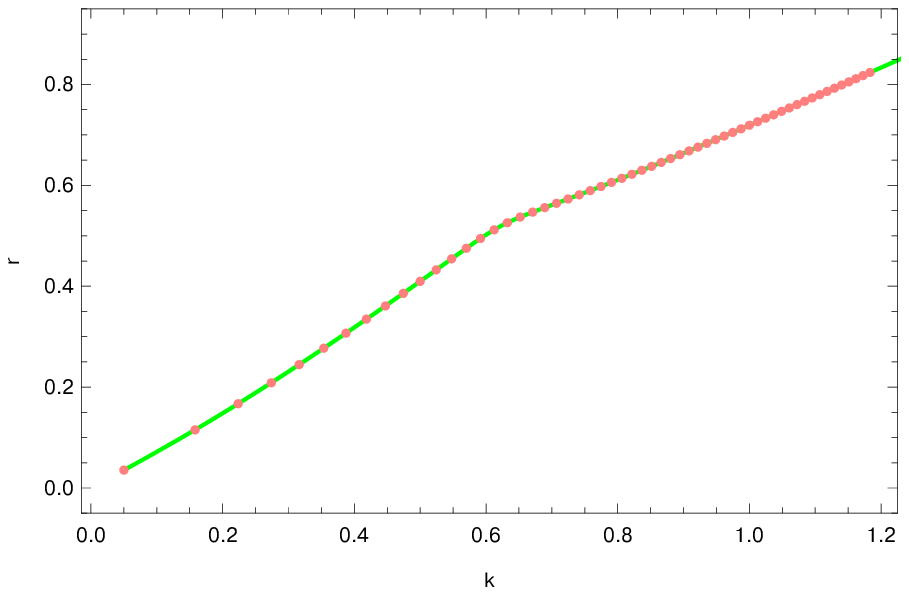}\,
\includegraphics[width=3in]{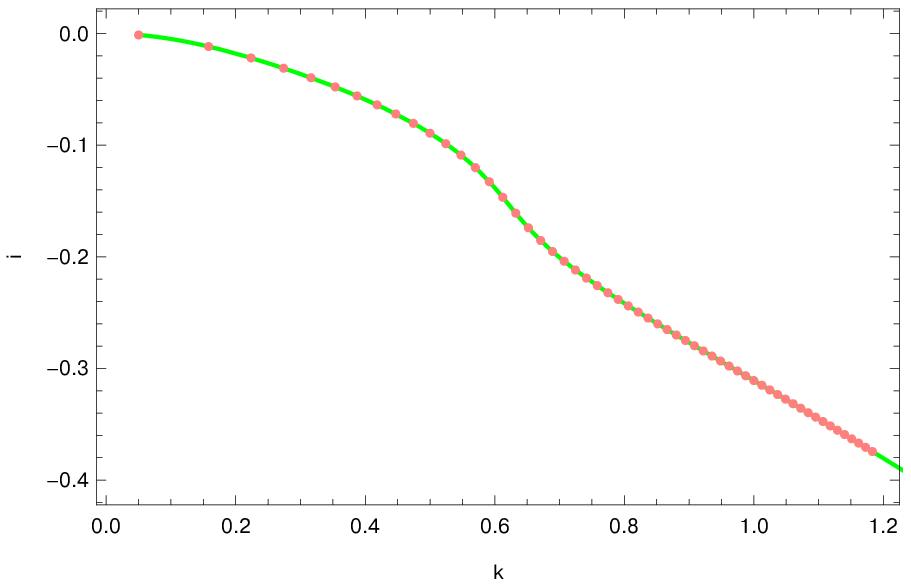}
\end{center}
  \caption{Comparison of the sound waves dispersion in the ordered phase at $b=-3$ extracted using the
  the $x$ radial coordinate (solid curves) and the $\hx$ radial coordinate, see \eqref{relatexxxhx},
  (the dots). 
} \label{figa2}
\end{figure}

Results reported in this paper involve numerical computation of the QNMs.
While some of the conclusions are robust and can be understood in the semi-analytic fashion,
\eg the existence of the unstable QNM in each  conformal order phase in the limit $b\to -\infty$
from the effective 1d Schr\"odinger problem, the bulk of the computations involves the heavy
numerics.  It is thus important to address the question of the numerical tests.
We summarize here the implicit and the explicit test.
\begin{itemize}
\item Thermal equilibrium phases of the holographic model \eqref{spsi} are conformal. This fact along
predicts the hydrodynamic transport, \ie the speed of the sound waves $c_s$ and the bulk viscosity
both in the symmetric and the symmetry broken phases. In fig.~\ref{fig3a} we presented deviations of these
quantities, obtained from the QNM computations,
from the expected values dictated by the conformal symmetry \eqref{cszeta}. 
\item The universality theorem of  \cite{Buchel:2003tz} implies that the shear viscosity
of the holographic conformal order is $s/(4\pi)$. This result was obtained evaluating the retarded
two-point correlation function of the boundary stress-energy tensor in a generic holographic model.
In fig.~\ref{fig3} we presented the deviation of the shear viscosity from the universal result obtained from the
dispersion relation of the QNMs in the shear channel.
\item In \cite{Buchel:2020thm} the holographic conformal order was constructed using the background metric
ansatz
\begin{equation}
ds_4^2=\frac{\alpha^2a(\hx)^2}{(2\hx-\hx^2)^{2/3}}\biggl(-(1-\hx)^2 dt^2+\left[dx_1^2+dx_2^2\right]\biggr)
+g_{\hx \hx} d\hx^2\,,
\eqlabel{oldmetric}
\end{equation}
where we denoted a radial coordinate as  $\hat{x}$  to distinguish it from the radial coordinate $x$ used here,
see \eqref{defx}. The two radial coordinates are related as
\begin{equation}
x=(2\hx -\hx^2)^{1/3}\,.
\eqlabel{relatexxxhx}
\end{equation}
Of course, the results should not depend whether we use $x$ or $\hx$ as a radial coordinate. We emphasize though
that the two computational frameworks are very different;  this is particularly profound in the
computation of the spectrum of the QNMs. Indeed, while the incoming-wave boundary condition for a typical gauge-invariant fluctuation $Z$ is $Z(x)\propto (1-x)^{-i\ww}$, the same boundary condition takes form
\begin{equation}
Z(\hx)\propto (1-\hx)^{-2i\ww}\,.
\eqlabel{bcnew}
\end{equation}
Since generically the dispersion relations $\ww(\kk)$ are complex, the $\Re-\Im$ part splits as in
\eqref{defsk} are different for the same normalizations of the wavefunctions (as in \eqref{soundk1}):
\begin{equation}
\Re[Z(x)]\ \ne \Re[Z(\hx)]\,,\qquad \Im[Z(x)]\ \ne \Im[Z(\hx)]\,.
\eqlabel{not}
\end{equation}
Of course, ultimately, this should not affect the computed spectrum $\ww(\kk)$.
In figs.~\ref{figa1} and \ref{figa2} we compare the spectra of the sound waves computed
using the $x$ radial coordinate (solid curves) and using the $\hx$ radial coordinate (dots)
in the holographic conformal order at $b=-5$ and $b=-3$ correspondingly. There is an excellent
agreement.
\item In fact, almost all computations presented in this paper were duplicated in $x$ and $\hx$
radial coordinates. For example, the fractional difference of the unstable QNM, the red dot in
fig.~\ref{fig6}, evaluated in two schemes is $\propto 10^{-9}$.
\end{itemize}

\section*{Acknowledgments}
Research at Perimeter
Institute is supported by the Government of Canada through Industry
Canada and by the Province of Ontario through the Ministry of
Research \& Innovation. This work was further supported by
NSERC through the Discovery Grants program.\bibliographystyle{JHEP}

\bibliography{cfttransport1}

\end{document}